\documentclass[10pt,preprint]{aastex}
\usepackage{relsize}
\shorttitle{Galaxy and AGN evolution in the infrared}
\shortauthors{Z.-Y. Cai et al.}

\begin{document}
\title{A hybrid model for the evolution of galaxies and Active Galactic Nuclei in the infrared}
\author
{Zhen-Yi~Cai\altaffilmark{1,2}, Andrea Lapi\altaffilmark{3,1}, Jun-Qing Xia\altaffilmark{4,1}, Gianfranco De Zotti\altaffilmark{1,5}, Mattia Negrello\altaffilmark{5}, Carlotta Gruppioni\altaffilmark{6}, Emma Rigby\altaffilmark{7},  Guillaume Castex\altaffilmark{8}, Jacques Delabrouille\altaffilmark{8},  Luigi Danese\altaffilmark{1}}
\altaffiltext{1}{Astrophysics Sector, SISSA, Via Bonomea 265, I-34136 Trieste, Italy; zcai@sissa.it}
\altaffiltext{2}{Department of Astronomy and Institute of Theoretical Physics and Astrophysics, Xiamen University, Xiamen 361005, P. R. China}
\altaffiltext{3}{Dipartimento di Fisica, Universit\`a `Tor Vergata', Via della Ricerca Scientifica 1, I-00133 Roma, Italy}
\altaffiltext{4}{Key Laboratory of Particle Astrophysics, Institute of High Energy Physics, Chinese Academy of Science, P.O.Box 918-3, Beijing 100049, P.R.China}
\altaffiltext{5}{INAF - Osservatorio Astronomico di Padova, Vicolo dell'Osservatorio 5, I-35122 Padova, Italy}
\altaffiltext{6}{INAF - Osservatorio Astronomico di Bologna, via Ranzani 1, I-40127 Bologna, Italy}
\altaffiltext{7}{Leiden Observatory, P.O. Box 9513, 2300 RA, Leiden, The Netherlands}
\altaffiltext{8}{APC, 10, rue Alice Domon et L\'eonie Duquet, 75205 Paris Cedex 13, France}

\def\lsim{\,\lower2truept\hbox{${<\atop\hbox{\raise4truept\hbox{$\sim$}}}$}\,}
\def\gsim{\,\lower2truept\hbox{${> \atop\hbox{\raise4truept\hbox{$\sim$}}}$}\,}

\begin{abstract}
We present a comprehensive investigation of the cosmological evolution of the luminosity function of galaxies and active galactic nuclei (AGN) in the infrared (IR). Based on the observed dichotomy in the ages of stellar populations of early-type galaxies on one side and late-type galaxies on the other, the models interprets the epoch-dependent luminosity functions at $z\ge 1.5$ using a physical model for the evolution of proto-spheroidal galaxies and of the associated AGNs, while IR galaxies at  $z< 1.5$ are interpreted as being mostly late-type ``cold'' (normal) and ``warm'' (starburst) galaxies. As for proto-spheroids, in addition to the epoch-dependent luminosity functions of stellar and AGN components separately, we have worked out, for the first time, the evolving luminosity functions of these objects as a whole (stellar plus AGN component), taking into account in a self-consistent way the variation with galactic age of the global SED. The model provides a physical explanation for the observed positive evolution of both galaxies and AGNs up to $z\simeq 2.5$ and for the negative evolution at higher redshifts, for the sharp transition from Euclidean to extremely steep counts at (sub-)mm wavelengths, as well as the (sub-)mm counts of strongly lensed galaxies, that are hard to account for by alternative, physical or phenomenological, approaches. The evolution of late-type galaxies and of $z<1.5$ AGNs is described using a parametric phenomenological approach.  The modeled AGN contributions to the counts and to the cosmic infrared background (CIB) are always subdominant. They are maximal at mid-IR wavelengths: the contribution to the 15 and $24\,\mu$m counts reaches 20\% above 10 and 2 mJy, respectively, while the contributions to the CIB are of 8.6\% and of 8.1\% at $15\,\mu$m and $24\,\mu$m, respectively. The model provides a good fit to the multi-wavelength (from the mid-IR to millimeter waves) data on luminosity functions at different redshifts and on number counts (both global and per redshift slices).  A prediction of the present model, useful to test it, is a systematic variation with wavelength of the populations dominating the counts and the contributions to the CIB intensity. This implies a specific trend for cross-wavelength CIB power spectra, that is found to be in good agreement with the data.
\end{abstract}



\section{Introduction}
\label{sec:intro}

The huge amount of infrared (IR) to millimeter-wave data that has been accumulating in the last several years have not yet led to a fully coherent, established picture of the cosmic star-formation history, of the IR evolution of Active Galactic Nuclei (AGNs), and of the inter-relations between star formation and nuclear activity.

Many, increasingly sophisticated, phenomenological models for the cosmological evolution of the galaxy and AGN luminosity functions over a broad wavelength range have been worked out \citep[e.g.][]{Bethermin2012a,Bethermin2011,Gruppioni2011,Rahmati2011,Marsden2011,Franceschini2010,
Valiante2009,LeBorgne2009,RowanRobinson2009}. These models generally include multiple galaxy populations, with different spectral energy distributions (SEDs) and different evolutionary properties, described by simple analytic formulae. In some cases also AGNs are taken into account. All of them, however, admittedly have limitations.

The complex combination of source properties (both in terms of the mixture of SEDs and of evolutionary properties), called for by the richness of data, results in a large number of parameters, implying substantial degeneracies that hamper the interpretation of the results. The lack of constraints coming from the understanding of the astrophysical processes controlling the evolution and the SEDs limits the predictive capabilities of these models. In fact, predictions of pre-{\it Herschel} phenomenological models, matching the data then available, yielded predictions for  {\it Herschel} counts quite discrepant from each other and with the data.

The final goal is a physical model linking the galaxy and AGN formation and evolution to primordial density perturbations. In this paper we make a step in this direction presenting a comprehensive `hybrid' approach, combining a physical, forward model for spheroidal galaxies and the early evolution of the associated AGNs with a phenomenological backward model for late-type galaxies and for the later AGN evolution. We start from the consideration of the observed dichotomy in the ages of stellar populations of early-type galaxies on one side and late-type galaxies on the other. Early-type galaxies and massive bulges of S$a$ galaxies are composed of relatively old stellar populations with mass-weighted ages of $\ga 8$--9 Gyr (corresponding to formation redshifts $z\ga 1$--1.5), while the disc components of spiral and irregular galaxies are characterized by significantly younger stellar populations. For instance, the luminosity-weighted age for most of S$b$ or later-type spirals is $\la 7$ Gyr (cf. Bernardi et al. 2010, their Fig.~10), corresponding to a formation redshift $z\la 1$. Thus proto-spheroidal galaxies are the dominant star-forming population at $z\ge 1.5$, while IR galaxies at  $z< 1.5$ are mostly late-type ``cold'' (normal) and ``warm'' (starburst) galaxies.

Fuller hierarchical galaxy formation models, whereby the mass assembly of galaxies is related to structure
formation in the dark matter and the star formation and merger histories of galaxies of all morphological types are calculated based on physical prescriptions have been recently presented by several groups \citep{Lacey2008,Fontanot2009,Narayanan2010,Shimizu2012}. However, the predictions for the IR evolution of galaxies are limited to a small set of wavelengths and frequently highlight serious difficulties with accounting for observational data \citep{Lacey2010,Niemi2012,Hayward2012}.

While the evolution of dark matter halos in the framework of the `concordance' $\Lambda$CDM cosmology is reasonably well understood thanks to N-body simulations such as the Millennium, the Millennium-XXL and the Bolshoi simulations \citep{Springel2005,Boylan-Kolchin2009,Angulo2012,Klypin2011}, establishing a clear connection between dark matter halos and visible objects proved to be quite challenging, especially at (sub-)mm wavelengths. The early predictions of the currently favoured scenario, whereby both the star-formation and the nuclear activity are driven by mergers, were more than one order of magnitude below the observed SCUBA $850\,\mu$m counts \citep{Kaviani2003,Baugh2005}. The basic problem is that the duration of the star-formation activity triggered by mergers is too short, requiring non standard assumptions either on the Initial Mass Function (IMF) or on dust properties to account for the measured source counts. The problem is more clearly illustrated in terms of redshift-dependent far-IR/sub-mm luminosity function, estimated on the basis of {\it Herschel} data \citep{Eales2010,Gruppioni2010,Lapi2011}. These estimates consistently show that $z\simeq 2$ galaxies with Star Formation Rates $\hbox{SFR}\simeq 300\,M_\odot\,\hbox{yr}^{-1}$ have comoving densities $\Phi_{300}\sim 10^{-4}\,\hbox{Mpc}^{-3}\,\hbox{dex}^{-1}$. The comoving density of the corresponding halos is $n(M_{\rm vir})\sim \Phi_{300}(t_{\rm exp}/\tau_{\rm SFR})$, where $M_{\rm vir}$ is the total virial mass (mostly dark matter), $\tau_{\rm SFR}$ is the lifetime of the star-forming phase and $t_{\rm exp}$ is the expansion timescale. For the fiducial lifetime $\tau_{\rm SFR}\simeq 0.7\,$Gyr advocated by \citet{Lapi2011}, $\log(M_{\rm vir}/M_\odot) \simeq 12.92$ while for $\tau_{\rm SFR}\simeq 0.1\,$Gyr, typical of a merger-driven starburst, $\log(M_{\rm vir}/M_\odot) \simeq 12.12$. Thus while the \citet{Lapi2011} model implies a $\hbox{SFR}/M_{\rm vir}$ ratio easily accounted for on the basis of standard IMFs and dust properties, the latter scenario requires a $\hbox{SFR}/M_{\rm vir}$ ratio more than a factor of 6 higher.

To reach the required values of $\hbox{SFR}/M_{\rm vir}$ or, equivalently, of $L_{\rm IR}/M_{\rm vir}$, \cite{Baugh2005} resorted to a top-heavy IMF while \citet{Kaviani2003} assumed that the bulk of the sub-mm emission comes from a huge amount of cool dust. But even tweaking with the IMF and with dust properties, fits of the sub-mm counts obtained within the merger-driven scenario \citep{Lacey2010,Niemi2012} are generally unsatisfactory. Further constraints on physical models come from the clustering properties of sub-mm galaxies that are determined by their effective halo masses. As shown by \citet{Xia2012} both the angular correlation function of detected sub-mm galaxies and the power spectrum of fluctuations of the cosmic infrared background indicate halo masses larger than implied by the major mergers plus top-heavy initial stellar mass function scenario \citep{Kim2011} and smaller than implied by cold flow models but consistent with the self-regulated baryon collapse scenario \citep{Granato2004,Lapi2006,Lapi2011}.

As is well known, the strongly negative K-correction emphasizes high-$z$ sources at (sub-)mm wavelengths. The data show that the steeply rising portion of the (sub-)mm counts is indeed dominated by ultra-luminous star-forming galaxies with a redshift distribution peaking at $z\simeq 2.5$ \citep{Chapman2005,Aretxaga2007,Yun2012,Smolcic2012}. As shown by \citet{Lapi2011}, the self-regulated baryon collapse scenario provides a good fit of the (sub-)mm data (counts, redshift-dependent luminosity functions) as well as of the stellar mass functions at  different redshifts. Moreover, the counts of strongly lensed galaxies were predicted with remarkable accuracy \citep{Negrello2007,Negrello2010,Lapi2012,Gonzalez-Nuevo2012}. Further considering that this scenario accounts for the clustering properties of sub-mm galaxies \citep{Xia2012}, we conclude that it is well grounded, and we adopt it for the present analysis. However, we upgrade this model in two respects. First, while, on one side, the model envisages a co-evolution of spheroidal galaxies and active nuclei at their centers, the emissions of the two components have been, so far, treated independently of each other. This is not a problem in the wavelength ranges where one of the two components dominates, as in the (sub-)mm region where the emission is dominated by star-formation, but is no longer adequate at mid-IR wavelengths, where the AGN contribution may be substantial. In this paper we present and exploit a consistent treatment of proto-spheroidal galaxies including both components. Second, while the steeply rising portion of (sub-)mm counts is fully accounted for by proto-spheroidal galaxies, late-type (normal and starburst) galaxies dominate both at brighter and fainter flux densities and over broad flux density ranges at mid-IR wavelengths. At these wavelengths, AGNs not associated to proto-spheroidal galaxies but either to evolved early type galaxies or to late-type galaxies are also important. Since we do not have a physical evolutionary model for late-type galaxies and the associated AGNs, these source populations have been dealt with adopting a phenomenological approach.

Another distinctive feature of the present model is that we have attempted to fit simultaneously the data over a broad wavelength range, from mid-IR to mm waves. As mentioned in several papers, this faces us with several challenges. First, the data come from different instruments and the relative calibration is sometimes problematic \citep[see the discussion in][]{Bethermin2011}. Systematic calibration offsets may hinder simultaneous fits of different data sets. For example, \citet{Marsden2011} pointed out that there is considerable tension between the SCUBA $850\,\mu$m counts and the AzTEC counts at 1.1\,mm, and indeed the $850\,\mu$m and mm-wave counts have been repeatedly corrected (generally downwards) as biases were discovered and better data were acquired. Also, the very complex SEDs in the mid-IR, where strong polycyclic aromatic hydrocarbon (PAH) emission features show up, make the counts exceedingly sensitive to the details of the spectral response function of the specific instrument and introduce large uncertainties in the conversion from broad-band measurements to monochromatic flux densities giving  rise to strong discrepancies among data sets nominally referring to the same wavelength. In fact, large discrepancies are  present among different determinations of $15\,\mu$m and $60\,\mu$m source counts.

The plan of the work is the following. In Section\,\ref{sect:protosph} we describe the physical model for the evolution of proto-spheroidal galaxies and of the associated AGNs and the SEDs adopted for these sources. Section~\ref{sect:late_type} deals with the evolutionary model for late-type galaxies and $z\le 1.5$ AGNs. In Section\,\ref{sect:formalism} we present the formalism to compute the source counts of unlensed and lensed sources, the cumulative flux density as a function of redshift and the contributions to the CIB. In Section\,\ref{sect:parameters} we report on the determination of the best fit values of the model parameters. In Section\,\ref{sect:results} the model results are compared with data on multi-frequency luminosity functions at various redshifts and on source counts, both total and per redshift slices. The multi-frequency power-spectra of CIB fluctuations implied by the model are discussed in Section\,\ref{sect:power_spectra}. Finally, Section\,\ref{sect:conclusions} contains a summary of the paper and our main conclusions.

Tabulations of multi-frequency model counts, redshift distributions, SEDs, redshift-dependent luminosity functions at several wavelengths, and a large set of figures comparing model predictions with the data are  available in the Web site http://people.sissa.it/$\sim$zcai/galaxy\_agn/.

Throughout this paper we adopt a flat cosmology with present day matter and baryon density, in units of the critical density, $\Omega_{\rm m,0}=0.27$ and $\Omega_{\rm b,0}=0.044$; Hubble constant $\hbox{h}=\hbox{H}_0/100=0.71$; spectrum of primordial density perturbations with slope $n=1$ and normalization on a scale of $8\,\hbox{h}^{-1}$ Mpc $\sigma_8=0.81$.

\section{Star-forming proto-spheroidal galaxies}\label{sect:protosph}

\subsection{Overview of the model}\label{sect:overview}

We adopt the model by \citet[][see also Lapi et al. 2006, 2011; Mao et al. 2007]{Granato2004} that interprets powerful high-$z$ sub-mm galaxies as massive proto-spheroidal galaxies in the process of forming most of their stellar mass. {It hinges upon high resolution numerical simulations showing that dark matter halos form in two stages \citep{Zhao2003,Wang2011,LapiCavaliere2011}. An early fast collapse of the halo bulk, including a few major merger events, reshuffles the gravitational potential and causes the dark matter and stellar components to undergo (incomplete) dynamical relaxation. A slow growth of the halo outskirts in the form of many minor mergers and diffuse accretion follows; this second stage has little effect on the inner potential well where the visible galaxy resides.}

{The star formation is triggered by the fast collapse/merger phase of the halo and is controlled by self-regulated baryonic processes. It is driven by the rapid cooling of the gas within a region with radius $\approx 30\%$ of the halo virial radius, i.e. of $\simeq 70(M_{\rm vir}/10^{13}\,M_\odot)^{1/3}[(1+z_{\rm vir})/3]^{-1}\,$kpc, where $M_{\rm vir}$ is the halo mass and $z_{\rm vir}$ is the virialization redshift, encompassing about $40\%$ of the total mass (dark matter plus baryons). The star formation and the growth of the central black-hole are regulated by the energy feedback from supernovae (SNe) and from the active nucleus, is very soon obscured by dust and is stopped by quasar feedback.  The AGN feedback is relevant especially in the most massive galaxies and is responsible for their shorter duration ($5-7\times 10^8\,$yr) of the active star-forming phase. In less massive proto-spheroidal galaxies the star formation rate is mostly regulated by SN feedback and continues for a few Gyr. Only a minor fraction of the gas initially associated to the dark matter halo is converted into stars. The rest is ejected by feedback processes.}

The equations governing the evolution of the baryonic matter in dark matter halos and the adopted values for the parameters are given in the Appendix where some examples of the evolution with galactic age (from the virialization time) of quantities related to the stellar and to the AGN component are also shown. For additional details and estimates of physically plausible ranges for each parameter we refer to \citet{Granato2004}, \citet{Lapi2006} and \citet{Mao2007}.  Since spheroidal galaxies are observed to be in passive evolution at $z\la 1-1.5$ \citep[e.g.][]{Renzini2006}, they are bright at sub-mm wavelengths only at higher redshifts.

\subsection{Luminosity functions}\label{sect:protoLF}

The bolometric luminosity function (LF) of proto-spheroids is obtained convolving the halo formation rate $dN_{\rm ST}(M_{\rm vir},z)/dt$ 
with the galaxy luminosity distribution, $P(L,z;M_{\rm vir})$. The halo formation rate is well approximated, for $z\gtrsim1.5$, by the positive term of the cosmic time derivative of the halo mass function $N_{\rm ST}$. For the latter, giving the average comoving number density of haloes of given mass, $M_{\rm vir}$, we adopt the \citet{ShethTormen1999} analytical expression
\begin{equation}
 N_{\rm ST}(M_{\rm vir}, z) dM_{\rm vir} = \frac{\bar\rho_{\rm m,0}}{M_{\rm vir}^2} f_{\rm ST}(\nu) \frac{d\ln\nu}{d\ln M_{\rm vir}}dM_{\rm vir},
\end{equation}
where $\bar\rho_{\rm m,0}=\Omega_{\rm m,0} \rho_{\rm c,0}$ is the present day mean comoving matter density of the universe and $\nu \equiv [\delta_{\rm c}(z)/\sigma(M_{\rm vir})]^2$, with $\delta_{\rm c}(z)=\delta_0(z)D(0)/D(z)$. The critical value of the initial overdensity that is required for spherical collapse at $z$, $\delta_0(z)$, is $\delta_{\rm c}(z)=\delta_0(z)D(0)/D(z)$ with \citep{NakamuraSuto1997}
$$\delta_0(z)=\frac{3(12\pi)^{2/3}}{20} [1+0.0123\log \Omega_{\rm m}(z)] \simeq 1.6865[1+0.0123\log \Omega_{\rm m}(z)].$$
The linear growth factor can be approximated as \citep{Lahav1991,Carroll1992}
$$D(z)=\frac{5\Omega_{\rm m}(z)}{2(1+z)}\Big/\left[\frac{1}{70} +\frac{209}{140}\Omega_{\rm m}(z) - \frac{1}{140}\Omega_{\rm m}^2(z) + \Omega_{\rm m}^{4/7}(z)\right].$$
The mass variance $\sigma(M_{\rm vir})$ of the primordial perturbation field smoothed on a scale containing a mass $M_{\rm vir}$ with a top-hat window function was computed using the \citet{Bardeen1986} power spectrum
with correction for baryons \citep{Sugiyama1995}, for our choice of cosmological parameters (see Section\,\ref{sec:intro}). The results are accurately approximated (error $<1\%$ over a broad range of $M_{\rm vir}$, $10^6< M_{\rm vir}/M_\odot < 10^{16}$) by
\begin{eqnarray}
\sigma(M_{\rm vir})\!\!\!\! &=&\!\!\!\! \frac{0.8}{0.84} \big[ 14.110393-1.1605397x-0.0022104939x^2\nonumber\\
\!\!\!\!&+&\!\!\!\!0.0013317476x^3 -2.1049631\times10^{-6}x^4 \big]
\end{eqnarray}
where $x \equiv \log (M_{\rm vir}/M_\odot)$. Furthermore
$$f_{\rm ST}(\nu) = A [1+(a\nu)^{-p}] \Big(\frac{a\nu}{2}\Big)^{1/2} \frac{e^{-a\nu/2}}{\pi^{1/2}},$$
where $A=0.322,\ p=0.3,\ {\rm and}\ a=0.707$.

The halo formation rate is then
\begin{eqnarray}
  & \mathlarger{{dN_{\rm ST}(M_{\rm vir}, z)\over dt}} = N_{\rm ST}(M_{\rm vir}, z) \mathlarger{{d\ln f_{\rm ST}(\nu)\over dt}}\nonumber\\
  &= - N_{\rm ST}(M_{\rm vir}, z) \mathlarger{\left[{a\delta_{\rm c}\over \sigma^2} + {2p\over \delta_{\rm c}} {\sigma^{2p}\over\sigma^{2p}+a^p \delta_{\rm c}^{2p}} - {1\over \delta_{\rm c}} \right] {d\delta_{\rm c}\over dz} {dz\over dt}} \nonumber\\
  &\simeq N_{\rm ST}(M_{\rm vir}, z) \mathlarger{\left[{a\nu\over 2} + {p\over 1+(a\nu)^{p}} \right] {d\ln \nu\over dz} \Big|{dz\over dt}\Big|},
\end{eqnarray}
where $dz/dt=- H_{0}(1+z)E(z)$ with $E(z) \equiv \sqrt{\Omega_{\Lambda,0} + \Omega_{\rm m,0}(1+z)^3}$.

The comoving differential luminosity function $\Phi(\log L, z)$, i.e. the number density of galaxies per unit $\log L$ interval at redshift $z$, is given by
\begin{eqnarray}\label{LF_scatter1}
 \Phi(\log L, z)\!\!\!\!  &=&\!\!\!\!  \int^{M^{\rm max}_{\rm vir}}_{M^{\rm min}_{\rm vir}}\!\!\!\!\!  dM_{\rm vir} \int^{z^{\rm max}_{\rm vir}}_z\!\!\!\!\!\! dz_{\rm vir} \Big|\frac{dt_{\rm vir}}{dz_{\rm vir}}\Big| \frac{dN_{\rm ST}}{dt_{\rm vir}}(M_{\rm vir}, z_{\rm vir})\cdot\nonumber\\
\!\!\!\!  &\cdot&\!\!\!\!    P(\log L,z; M_{\rm vir}, z_{\rm vir}),
\end{eqnarray}
where $P(\log L, z; M_{\rm vir}, z_{\rm vir})$ is the luminosity distribution of galaxies at redshift $z$ inside a halo of mass $M_{\rm vir}$ virialized at redshift $z_{\rm vir}$. We set $z^{\rm min}_{\rm vir}=1.5$ and $z^{\rm max}_{\rm vir}=12$.

As mentioned in Section\,\ref{sect:overview} the total luminosity of a galaxy is the sum of those of the stellar component and of the active nucleus. For each component we assume a log-normal luminosity distribution
\begin{equation}
	P[\log L|\log \bar L] d\log L = \frac{\exp [-\log^2(L/\bar L)/2\sigma^2]}{\sqrt{2\pi\sigma^2}} d\log L,
\end{equation}
with dispersion $\sigma_* = 0.10$ around the mean stellar luminosity $\bar L_*(z; M_{\rm vir}, z_{\rm vir})$ and $\sigma_\bullet = 0.35$ around the mean AGN luminosity $\bar L_\bullet(z; M_{\rm vir}, z_{\rm vir})$. The mean luminosities are computed solving the equations detailed in the Appendix. The higher luminosity dispersion for the AGN component reflects its less direct relationship, compared to the stellar component, with $M_{\rm vir}$ and $z_{\rm vir}$. The distribution of the total luminosity, $L_{\rm tot} = L_* + L_\bullet$, is then \citep{Dufresne2004}
\begin{eqnarray}\label{LF_scatter2}
	&P[\log L_{\rm tot}|\log \bar L_*, \log \bar L_\bullet] d\log L_{\rm tot} = d\log L_{\rm tot} \nonumber\\
	& \times \mathlarger{\int^{\log L_{\rm tot}}_{-\infty} \frac{dx}{2\pi \sigma_* \sigma_\bullet} \frac{L_{\rm tot}}{L_{\rm tot} - 10^x}} \exp \{-(x - \log \bar L_*)^2/2\sigma^2_* \} \nonumber\\
	& \ \ \ \ \ \ \ \ \ \times \exp \{-[\log (L_{\rm tot} - 10^x) - \log \bar L_\bullet]^2/2\sigma^2_\bullet\}.
\end{eqnarray}
In the upper left panel of Fig.~\ref{fig:sph_LFbol} we show, as an example, the bolometric luminosity functions at $z=1.5$ of the stellar and of the AGN components, as well as the luminosity function of the objects as a whole. As shown by eq.~(\ref{LF_scatter2}) the latter is different from the sum of the first two, although in this case the difference is difficult to perceive. The bright end is dominated by QSOs shining unobstructed after having swept away the interstellar medium of the host galaxy. In this phase the QSOs reach their maximum luminosity. Around $\log (L_{\rm bol}/L_\odot) \simeq 13$ the AGNs and the starbursts give similar contributions to the bolometric luminosity function. The inflection at $\log(L_{\rm IR}/L_\odot)\simeq 11.7$ corresponds to the transition from the regime where the feedback is dominated by supernovae (lower halo masses) to the regime where it is dominated by AGNs. While the star formation in massive halos is abruptly stopped by the AGN feedback after 0.5--$0.7\,$Gyr, it lasts much longer in smaller galaxies, implying a fast increase of their number density.  

The upper right panel of the same figure illustrates the evolution with cosmic time of the global luminosity function. The cooling and free-fall timescales shorten with increasing redshift because of the increase of the matter density and this drives a positive luminosity evolution, thwarted by the decrease in the comoving density of massive halos. The two competing factors result, for both the starburst and the AGN component (see the lower panels of the figure), in a positive evolution up to $z\simeq 2.5$ followed by a decline at higher $z$, consistent with the observational determinations by \citet{Gruppioni2010} and \citet{Lapi2011} for the starburst component and by \citet{Assef2011} and \citet{Brown2006} for AGNs. The decrease of the luminosity function at low luminosities, more clearly visible at the higher redshifts, is an artifact due to the adopted lower limit for the considered halo masses. This part of the luminosity function however does not contribute significantly to the observed statistics and therefore is essentially irrelevant here.  Below the minimum virialization redshift, $z^{\rm min}_{\rm vir}=1.5$, the bolometric luminosity function of proto-spheroidal galaxies rapidly declines as they evolve towards the `passive' phase. The decline is faster at the bright end (above $\log (L_{\rm bol}/L_\odot) \simeq 12$) since the switching off of the star formation for the more massive halos occurs on a shorter timescale.

The monochromatic luminosity functions of each component or of objects as a whole can obviously be computed using the same formalism, given the respective spectral energy distributions (SEDs). We define $\bar {\cal L}_{*, \nu} \equiv \nu \bar{L}_{*, \nu} = \nu f_*(\nu) \bar{L}_{*, {\rm IR}}$, $\bar{\cal L}_{\bullet, \nu} \equiv \nu \bar{L}_{\bullet, \nu} = \nu f_\bullet(\nu) \bar{L}_{\bullet, {\rm bol}}$, and $\bar {\cal L}_{\nu} \equiv \bar{\cal L}_{*, \nu}  + \bar{\cal L}_{\bullet, \nu}$, where $f(\nu)$ is the normalized SED ($\int\! d\nu\,f(\nu)=1$).

Since the model cannot follow in detail the evolution of the AGN SEDs during the short phase when they shine unobstructed by the interstellar medium of the host galaxy, the distinction between obscured and unobscured AGNs in the model is made in two ways. First, following \citet{Lapi2006}, we choose a fixed optical (B-band) ``visibility time'',   $\Delta t_{\rm vis}=5\times 10^7\,$yr, consistent with current estimates of the optically bright QSO phase. Alternatively, we set the beginning of the optical bright phase at the moment when the gas mass fraction is low enough to yield a low optical depth. We estimate that this corresponds to a gas fraction within the dark matter potential well $f_{\rm gas}=M_{\rm gas}/M_{\rm vir} \lesssim f_{\rm gas,crit}=0.03$. The two approaches give very similar results and we have chosen the criterion $f_{\rm gas} \lesssim f_{\rm gas,crit}$ to compute the luminosity functions at optical wavelengths.



\subsection{Spectral energy distributions}\label{sect:sed_proto}

Although there is evidence that the galaxy SEDs vary with luminosity \citep[e.g.][]{Smith2012}, \citet{Lapi2011} have shown that the sub-mm data can be accurately reproduced using a single SED for proto-spheroidal galaxies, i.e. the SED of the strongly lensed $z\simeq 2.3$ galaxy SMM~J2135-0102 \citep{Swinbank2010,Ivison2010}, modeled using GRASIL \citep{Silva1998}. The basic reason for the higher uniformity of the SEDs of high-$z$ active star-forming galaxies compared to galaxies at low-$z$ is that the far-IR emission of the former objects comes almost entirely from dust in molecular clouds, heated by newly formed stars, while in low-$z$ galaxies there are important additional contributions from colder `cirrus' heated by older stellar populations.

This SED worked very well at sub-mm wavelengths but yielded mm-wave counts in excess of the observed ones. To overcome this problem the sub-mm slope of the SED has been made somewhat steeper, preserving the consistency with the photometric data on SMM~J2135-0102 (see Fig.~\ref{fig:seds_sph}). Moreover, the SED used by Lapi et al. has a ratio between the total (8--$1000\,\mu$m) IR and the $8\,\mu$m luminosity ($\hbox{IR}8=L_{\rm IR}/L_8$) of $\simeq 30$, far higher than the mean value for $z\simeq 2$ galaxies \citep[$\hbox{IR}8\simeq 9$,][]{Reddy2012}. We have therefore modified the near- and mid-IR portions of the SED adopting a shape similar to that of Arp~220. The contribution of the passive evolution phase of early-type galaxies is small in the frequency range of interest here and will be neglected.

As mentioned in Section\,\ref{sect:overview}, the model follows the AGN evolution through two phases (a third phase, reactivation, will be considered in Section\,\ref{sect:reactiv_AGN}). For the first phase, when the black-hole growth is enshrouded by the abundant, dusty interstellar medium (ISM) of the host galaxy, we adopt the SED of a heavily absorbed AGN taken from the AGN SED library by \citet{GranatoDanese1994}. Note that these objects differ from the classical type-2 AGNs because they are not obscured by a circum-nuclear torus but by the more widely distributed dust in the host galaxy. They will be referred to as type-3 AGNs. In the second phase the AGN shines after having swept out the galaxy ISM. For this phase we adopted the mean QSO SED by \citet{Richards2006} extended to sub-mm wavelengths assuming a grey-body emission with dust temperature $T_{\rm dust} = 80\, \rm K$ and emissivity index $\beta = 1.8$. These SEDs imply that the IR (8--$1000\,\mu$m) band comprises 92\% of the bolometric luminosity of obscured AGNs and 19\% of that of the unobscured ones. As illustrated by Fig.~\ref{fig:seds_sph}, except in the rare cases in which the AGN bolometric luminosity is much larger than that of the starburst, the AGN contribution is small at (sub-)mm wavelengths, while it is important and may be dominant, in the mid-IR. This implies that the statistics discussed here are insensitive to the parameters describing the extrapolation of the Richards et al. SED to (sub-)mm wavelengths.

Figure~\ref{fig:seds_sph_tM} shows the global SEDs and the contributions of the stellar and AGN components for 2 galaxy ages and three host halo masses virialized at $z_{\rm vir} = 3$. The shorter evolution timescale of the AGNs is clearly visible. It is worth noticing that the effect of feedback as a function of halo mass on the SFR is very different from that on accretion onto the supermassive black-hole. In the less massive halos the AGN feedback has only a moderate effect on the evolution of the SFR and of the accretion rate, that are mostly controlled by the SN feedback. With reference to the figure, for $\log(M_{\rm vir}/M_\odot)=11.4$, the star-formation continues at an almost constant rate for a few Gyrs. On the other hand the accretion rate onto the central black-hole is at the Eddington limit only up to an age of $\simeq 0.3\,$Gyr and afterwards drops to a strongly sub-Eddington regime. This is because the growth rate of the reservoir is approximately proportional to the SFR (and therefore slowly varying for few Gyrs) while the accretion rate grows exponentially until the mass contained in the reservoir is exhausted. From this moment on the accretion rate is essentially equal to the (strongly sub-Eddington) inflow rate.  For more massive halos the quenching of both the SFR and of the accretion occurs more or less simultaneously at ages of $\simeq 0.5$--0.6\,Gyr, but while the SFR stops very rapidly, the AGN activity continues until the flow of the matter accumulated in the reservoir runs out. At ages $\gsim 0.6$\,Gyr the more massive galaxies are in passive evolution and therefore very weak in the far-IR while star-formation and the dust emission are still present in lower-mass galaxies.

\section{Low redshift ($z \lesssim 1.5$) populations}\label{sect:late_type}

\subsection{Late-type and starburst galaxies}\label{sect:late_type_sfgs}

We consider two $z \lesssim 1.5$ galaxy populations: ``warm" starburst galaxies and ``cold" (normal) late-type galaxies. For the IR luminosity function of both populations we adopt the functional form advocated by \citet{Saunders1990}:
\begin{equation}\label{eq:LF}
	\Phi (\log L_{\rm IR}, z) d\log L_{\rm IR} = \Phi^* \Big(\frac{L_{\rm IR}}{L^*}\Big)^{1-\alpha}
	 \times \exp \Big[- \frac{\log ^2 (1+L_{\rm IR}/L^*)}{2\sigma^2} \Big] d\log L_{\rm IR}
\end{equation}
where the characteristic density $\Phi^*$ and luminosity $L^*$, the low-luminosity slope $\alpha$ and the dispersion $\sigma$ of each population are, in principle, free parameters. However, the low-luminosity portion of the luminosity function is dominated by ``cold'' late-type galaxies and, as a consequence, the value of $\alpha$ of the warm population is largely unconstrained; we have fixed it at $\alpha_{\rm warm}=0.01$. In turn, the ``warm'' population dominates at high luminosities so that the data only imply an upper limit to $\sigma_{\rm cold}$. We have set $\sigma_{\rm cold}=0.3$.

For the ``warm'' population we have assumed power law density and luminosity evolution [$\Phi^*(z) = \Phi^*_0(1+z)^{\alpha_\Phi}$; $L^*(z)=L^*_0(1+z)^{\alpha_{\rm L}}$] up to $z_{\rm break}=1$, $\alpha_\Phi$ and $\alpha_{\rm L}$ being free parameters. The ``cold'' population comprises normal disc galaxies for which chemo/spectrophotometric evolution models \citep{Mazzei1992,Colavitti2008} indicate a mild (a factor $\simeq 2$ from $z=0$ to $z=1$) increase in the star formation rate, hence of IR luminosity, with look-back time. Based on these results we take, for this population, $\alpha_{\rm L}=1$ and no density evolution. At $z>z_{\rm break}$ both $\Phi^*(z)$ and $L^*(z)$ are kept to the values at $z_{\rm break}$ multiplied by the smooth cut-off function $\{1 - {\rm erf}[(z-z_{\rm cutoff})/\Delta z)] \}/2$, with $z_{\rm cutoff}=2$ and $\Delta z = 0.5$. The choice of the redshift cutoff for both populations of late-type galaxies is motivated by the fact that the disc component of spirals and the irregular galaxies are characterized by relatively young stellar populations (formation redshift $z\la 1$--1.5). Above $z=1.5$ proto-spheroidal galaxies (including bulges of disk galaxies) dominate the contribution to the luminosity function, at least in the observationally constrained luminosity range. The other parameters are determined by minimum $\chi^2$ fits to selected data sets, as described in Sect.~\ref{sect:parameters}. Their best fit values and the associated uncertainties are listed in Table~\ref{tab:parameters}.

Although there is clear evidence of systematic variations of the IR SEDs of low-$z$ galaxies with luminosity \citep[e.g.,][]{Smith2012}, we tried to fit the data with just 2 SEDs, one for the ``warm'' and one for the ``cold'' population. These SEDs were generated by combining those of \citet{DaleHelou2002}, that are best determined at mid-IR wavelengths, with those of \citet{Smith2012}, primarily based on \textit{Herschel} data in the range 100-500 $\mu$m. \citet{DaleHelou2002} give SED templates for several values of the 60 to 100 $\mu$m flux density ratio, $\log[f_\nu(60 \mu m)/f_\nu(100 \mu m)]$. Using the relation between this ratio and the 3 to 1100 $\mu$m luminosity, $L_{\rm TIR}$, given by \citet{Chapman2003} we established a correspondence between their SEDs and those by Smith et al., labeled by the values of $\log(L_{\rm IR}/L_\odot)$. The combined SEDs are based on Smith et al. above $100\,\mu$m and on Dale \& Helou at shorter wavelengths. By trial and error we found that the best fit to the data is obtained using for the ``cold'' population the SED corresponding to $\log(L_{\rm IR}/L_\odot) = 9.75$ (actually the SEDs change very slightly for $\log(L_{\rm IR}/L_\odot) \lesssim 9.75$) and for the ``warm'' population the SED corresponding to $\log(L_{\rm IR}/L_\odot) = 11.25$. These 2 SEDs are displayed in Fig.~\ref{fig:seds_sfg}.


\subsection{Reactivated AGNs}\label{sect:reactiv_AGN}

In the framework of our reference galaxy and AGN evolutionary scenario, most of the growth of super-massive black holes is associated to the star forming phase of spheroidal components of galaxies at $z\gtrsim 1.5$ when the great abundance of interstellar medium favours high accretion rates, at, or even slightly above, the Eddington limit. At later cosmic times the nuclei can be reactivated by, e.g., interactions, mergers or dynamical instabilities. The accretion rates are generally strongly sub-Eddington. Our evolutionary scenario cannot predict their amplitudes and duty cycles. We therefore adopted, also for these objects, a phenomenological backward evolution model analogous to that used for the ``warm'' galaxy population, i.e. luminosity functions of the same form of eq.~(\ref{eq:LF}) and power-law density and luminosity evolution with the same break and cutoff redshifts. However the parameters of the luminosity functions refer to $12\,\mu$m (see Sect.~\ref{sect:late_type_sfgs} and Table~\ref{tab:parameters}). The data do not allow a determination of the slopes, $\alpha$, of the faint portions of the luminosity functions. We have set $\alpha=1.1$ for type-1 AGNs and $\alpha=1.5$ for type-2. The steeper slope for type-2 was chosen on account of the fact that these dominate over type-1 at low luminosities. {As in the case of normal late-type and of starburst galaxies, the other parameters are obtained by minimum $\chi^2$ fits, as detailed in Sect.~\ref{sect:parameters}, and the best fit values are listed, with their uncertainties, in Table~\ref{tab:parameters}. For type-2 AGNs pure density evolution was found to be sufficient to account for the data.

For type-1 AGNs we adopted the mean QSO SED by \citet{Richards2006}, extended to mm wavelengths as described in Sect.\,\ref{sect:sed_proto}, while for type-2 AGNs we adopted the SED of the local AGN dominated ULIRG Mrk 231, taken from the SWIRE library \citep{Polletta2007}. These SEDs are shown in Fig.~\ref{fig:seds_agn} where the SED of type-3 AGNs associated to dusty star-forming proto-spheroidal galaxies is also plotted for comparison. The SED of type-3 AGNs is the most obscured at optical/near-IR wavelengths due to the effect of the dense, dusty interstellar medium of the high-$z$ host galaxies. This means that the counts at optical/near-IR wavelengths are dominated by type-1 AGNs with type-2 AGNs becoming increasingly important in the mid-IR. The 3 AGN populations have approximately the same ratio between the rest-frame $12\,\mu$m and the bolometric luminosity, as first pointed out by \citet{SpinoglioMalkan1989}.

The type-1/type-2 space density ratio yielded by the model increases with luminosity, consistent with observations \citep[e.g,][]{Burlon2011} and with the receding torus model \citep{Lawrence1991}. In the framework of the standard unified model of AGNs type-1 and type-2 AGNs differ only in terms of the angle which the observer's line of sight makes with the axis of a dusty torus. If the line of sight to the central region is blocked by the torus, the AGN is seen as a type-2. According to the receding torus model the opening angle of the torus (measured from the torus axis to the equatorial plane) is larger in more luminous objects, implying that obscuration is less common in more luminous AGNs. Since our model implies that type-1 AGNs (but not type-2's) are evolving in luminosity, they become increasingly dominant with increasing redshift.

\section{Source counts and contributions to the background}\label{sect:formalism}

The surface density of sources per unit flux density and redshift interval is
\begin{equation}
	\frac{d^3N(S_\nu, z)}{dS_\nu dz d\Omega}= \frac{\Phi(\log {L}_{\nu'}, z)}{L_{\nu'}\ln10} \frac{dL_{\nu'}}{dS_\nu} \frac{d^2V}{dzd\Omega}\ ,
\end{equation}
where $\nu' = \nu (1+z)$,
\begin{equation}
	S_\nu = \frac{(1+z)L_{\nu'}}{4\pi D^2_{\rm L}(z)},
\end{equation}
the comoving volume per unit solid angle is
\begin{equation}
	\frac{d^2V}{dzd\Omega} = \frac{c}{H_0} \frac{(1+z)^2D^2_{\rm A}(z)}{E(z)},
\end{equation}
and the luminosity distance $D_{\rm L}$ and the angular diameter distance $D_{\rm A}$ are related, in a flat universe, by
\begin{equation}
	\frac{D_{\rm L}}{1+z} = (1+z) D_{\rm A} = \frac{c}{H_0} \int^z_0 \frac{dz'}{E(z')}.
\end{equation}
The differential number counts, i.e., the number of galaxies with flux density in the interval $S_\nu \pm dS_\nu/2$ at an observed frequency $\nu$ per unit solid angle, are then
\begin{equation}
	\frac{d^2N}{dS_\nu d\Omega}(S_\nu) = \int^{z_{\rm max}}_{z_{\rm min}} dz \frac{\Phi(\log {L}_{\nu'}, z)}{L_{\nu'}\ln10} \frac{dL_{\nu'}}{dS_\nu} \frac{d^2V}{dzd\Omega}\ .
\end{equation}
The integral number counts, i.e., the number of galaxies with flux density $S_\nu>S_{\nu, \rm inf}$ at frequency $\nu$ per unit solid angle, are given by
\begin{equation}
	\frac{dN}{d\Omega}(S_\nu > S_{\nu,\rm inf}) = \int^{z_{\rm max}}_{z_{\rm min}} dz \frac{d^2V}{dzd\Omega}
	 \int^{\infty}_{\log { L}_{\nu',\rm inf}} \Phi(\log {L}_{\nu'}, z) d\log { L}_{\nu'},
\end{equation}
where $\nu'=(1+z)\nu$ and $L_{\nu',\rm inf}$ is the monochromatic luminosity of a source at the redshift $z$ observed to have a flux density $S_{\nu,\rm inf}$. Counts (per steradian) dominated by local objects ($z\ll 1$) can be approximated as
\begin{equation}
S^{2.5}_{\nu} \frac{d^2N}{dS_\nu d\Omega} \simeq \frac{1}{4\pi} \frac{1}{4\sqrt{\pi}} \int^{\infty}_{0} \Phi(\log {L}_\nu, z \simeq 0) L_\nu^{3/2} d\log {L}_\nu\ .
\end{equation}
The redshift distribution, i.e. the surface density of sources with observed flux densities greater than a chosen limit $S_{\nu, \rm inf}$ per unit redshift interval, is
\begin{equation}
	\frac{d^2N}{dzd\Omega}(z, S_\nu > S_{\nu, \rm inf}) = \int^{\infty}_{S_{\nu, \rm inf}} \frac{d^3N}{dS^\prime_{\nu}dzd\Omega} dS^\prime_{\nu}\ .
\end{equation}
The steepness of the (sub-)mm counts of proto-spheroidal galaxies and their substantial redshifts imply that their counts are strongly affected by the magnification bias due to gravitational lensing \citep{Blain1996,Perrotta2002,Perrotta2003,Negrello2007}:
%
\begin{equation}\label{eq:lensed_counts}
	\frac{d^3N_{\rm lensed}(S_\nu, z)}{d\log S_{\nu} dz d\Omega}\!\!  = \!\!\int_{\mu}\!\! d\mu
	  \frac{d^3N(S_\nu/\mu, z)}{d\log S_\nu dz d\Omega} \frac{dP(\mu|z)}{d\mu}\ ,
\end{equation}
where $dP/d\mu$ is the amplification distribution that describes the probability for a source at redshift $z$ to be amplified by factor $\mu$. Here we have approximated to unity the factor $1/\langle\mu\rangle$ that would have appeared on the right-hand side, as appropriate for large-area surveys \citep[see][]{JainLima2011}.

We have computed $dP/d\mu$ using the SISSA model worked out by \citet{Lapi2012}. The differential counts including the effect of lensing can be computed integrating eq.~(\ref{eq:lensed_counts}) over $z$. The effect of lensing on counts of other source populations and on proto-spheroidal counts at shorter wavelengths is small and will be neglected in the following.

Interesting constraints on the halo masses of proto-spheroidal galaxies come from the auto- and cross-correlation functions of intensity fluctuations. A key quantity in this respect is the flux function, $d^2S_\nu/dzd\Omega$, i.e. the redshift distribution of the cumulative flux density of sources below the detection limit $S_{\nu, \rm lim}$
\begin{equation}
	\frac{d^2S_\nu}{dzd\Omega} = \int^{S_{\nu, \rm lim}}_{0} \frac{d^3N}{dS_\nu dzd\Omega} S_\nu dS_\nu\ .
\end{equation}
%

The contribution of a source population to the extragalactic background at the frequency $\nu$ is
\begin{equation}
	I_\nu = \int^\infty_0 S_\nu \frac{d^2N(S_\nu)}{dS_\nu d\Omega} dS_\nu\ .
\end{equation}

\section{Determination of the best fit values of the parameters}\label{sect:parameters}

A minimum $\chi^2$ approach for estimating the optimum values of the parameters of the physical model for proto-spheroidal galaxies and associated AGNs is unfeasible because of the lengthy calculations required. Some small adjustments compared to earlier versions \citep{Granato2004,Lapi2006,Mao2007} were made, by trial and error, to improve the agreement with observational estimates of luminosity functions at $z>1.5$. An outline of the model, including the definition of the relevant parameters, is presented in \appendixname\,\ref{sect:appendix}. The chosen values are listed in Table~\ref{tab:parameters_proto}. Discussions of physically plausible ranges can be found in \citet{Granato2004}, \citet{Cirasuolo2005}, \citet{Lapi2006}, \citet{Shankar2006}, \citet{Cook2009} and \citet{Fan2010}.

On the contrary, the minimum $\chi^2$ approach was applied to late-type/starburst galaxies and to reactivated AGNs. The $\chi^2$ minimization was performed using the routine \textit{MPFIT}\footnote{http://purl.com/net/mpfit} exploiting the Levenberg-Marquardt least-squares method \citep{More1978,Markwardt2009}.

The huge amount of observational data in the frequency range of interest here and the large number of parameters coming into play forced us to deal with subsets of parameters at a time using specific data for each subset. The parameters of the evolving AGN luminosity functions were obtained using:
\begin{itemize}
	\item the \textit{B}-band local QSO luminosity function of \citet{HartwickSchade1990},
	\item the \textit{g}-band QSO luminosity functions at $z=0.55$ and $0.85$ of \citet{Croom2009},
	\item the $z \lesssim 0.75$, 1.24 $\mu$m AGN luminosity functions of \citet{Assef2011},
	\item the bright end [$\log(L_{60}/L_\odot) \geq 12$] of the local 60 $\mu$m luminosity function  of \citet{Takeuchi2003},
	\item the \textit{Spitzer} AGN counts at 8 and $24\,\mu$m of \citet{Treister2006}.
\end{itemize}
The \textit{B}- and \textit{g}-band luminosity functions were used to constrain the parameters of type-1 AGNs (type-2 being important only at the low luminosity end) while the 1.24 $\mu$m luminosity functions were regarded as made by a combination of type-1 and type-2 AGNs, the latter being dominant at low luminosities.

As for the evolving luminosity functions of ``warm'' and ``cold'' galaxy populations we used the following data sets:
\begin{itemize}
	\item the \textit{IRAS} $60\,\mu$m local luminosity function of \citet{SoiferNeugebauer1991},
	\item the \textit{Planck} 350, 550, and $850\,\mu$m local luminosity functions of \citet{Negrello2012},
	\item the \textit{Spitzer} MIPS counts at 24, 70, and $160\,\mu$m of \citet{Bethermin2010},
	\item the \textit{Herschel} PACS counts at $160\,\mu$m of \citet{Berta2011},
	\item the \textit{Herschel} SPIRE counts at 250, 350, and $500\,\mu$m \citep{Bethermin2012b}.
\end{itemize}
The fits of the counts were made after having subtracted the contributions of proto-spheroidal galaxies, which are only important at wavelengths $\ge 160\,\mu$m. The best-fit values of the parameters are listed in Table~\ref{tab:parameters}, where values without errors denote parameters that were kept fixed, as mentioned in Sect.\,\ref{sect:late_type}. 

In comparing model results with observational data the instrumental spectral responses were taken into account. This is especially important in the mid-IR because of the complexity of the SEDs due to PAH emission lines. The monochromatic luminosity at the effective frequency $\nu_{\rm eff}$ in the observer's frame is given by:
\begin{equation}\label{eq:SM_SED}
L(\nu_{\rm eff}) \equiv \int T(\nu') L_{\nu'(1+z)} d\nu' \Big/ \int T(\nu') d\nu'
\end{equation}
where $T(\nu)$ is spectral response function and the integration is carried out over the instrumental band-pass. When the model is compared with luminosity function data at frequency $\nu_i$ (in the source frame) coming from different instruments for sources at redshift $z$ we use the response function of the instrument for which $\nu_{\rm eff}$ is closest to $\nu_i/(1+z)$. In the case of source counts we use the response function appropriate for the most accurate data.

\section{Results}\label{sect:results}

\subsection{Model versus observed luminosity functions and redshift distributions}\label{sect:results_lf_rd}

The most direct predictions of the { {\it physical model} for proto-spheroidal galaxies are the redshift-dependent SFRs and accretion rates onto the super-massive black-holes as a function of halo mass. During the dust enshrouded evolutionary phase the SFRs can be immediately translated into the IR (8--$1000\,\mu$m) luminosity functions of galaxies. As mentioned above, according to our model, the transition from the dust-obscured to the passive evolution phase is almost instantaneous and we neglect the contribution of passive galaxies to the IR luminosity functions. In turn, the accretion rates translate into bolometric luminosities of AGNs given the mass-to-light conversion efficiency for which we adopt the standard value $\epsilon =0.1$. The SEDs then allow us to compute the galaxy and AGN luminosity functions at any wavelength.}

{In contrast, the {\it phenomenological model} for late-type/starburst galaxies yields directly the redshift-dependent IR luminosity functions and that for reactivated AGNs yields the $12\,\mu$m  luminosity functions. Again these can be translated to any wavelength using the SEDs described in the previous sections.}

In Fig.~\ref{fig:LF_IR} the model IR luminosity functions are compared with observation-based determinations at different redshifts. At $z>1.5$ the dominant contributions come from the stellar and AGN components of proto-spheroidal galaxies. These contributions fade at lower redshifts and essentially disappear at $z<1$. The model implies that AGNs associated to proto-spheroidal galaxies are important only at luminosities higher than those covered by the \citet{Lapi2011} luminosity functions which therefore have been converted to bolometric luminosity functions using their galaxy SED, i.e. neglecting the AGN contribution, so that $\log (L_{\rm IR}/L_\odot) = \log(L_{100}/L_\odot) + 0.21$ and $\log (L_{\rm IR}/L_\odot) = \log(L_{250}/L_\odot) + 1.24$. At $z\le 1.5$  ``warm'' and ``cold'' star forming galaxies take over, ``cold'' galaxies being important only at low luminosities. Type-2 AGNs (long-dashed pink lines) may dominate at the highest IR luminosities while type-1 AGNs (long-dashed light-blue lines) are always sub-dominant (in the IR).

The scale on the top $x$-axis in Fig.~\ref{fig:LF_IR} gives the star formation rates corresponding to the IR luminosities
\begin{equation}
\log \Big(\frac{L_{\rm IR}}{L_\odot}\Big) = \log \Big(\frac{{\rm SFR}}{M_\odot\ {\rm yr}^{-1}}\Big) + 9.892,
\end{equation}
and is therefore meaningful only to the extent that the AGN contribution is negligible. Moreover, the normalization constant applies to high-$z$ proto-spheroidal galaxies whose IR luminosity comes almost entirely from star-forming regions. For more evolved galaxies older stellar populations can contribute significantly to the dust heating \citep{daCunha2012}; therefore $L_{\rm IR}$ is no longer a direct measure of the star formation rate and therefore the upper scale has to be taken as purely indicative.

Observational determinations of luminosity functions are available in many wave bands and for many cosmic epochs. The comparison between the model and the observed \textit{g}-band ($0.467\,\mu$m) AGN luminosity functions at several redshifts is presented in Fig.~\ref{fig:LFnu_g}, while the comparison in the \textit{J}-band ($1.24\,\mu$m) is shown in Fig.~\ref{fig:LFnu_J}. {The conversion from monochromatic absolute AB magnitude $M_{\lambda, \rm AB}$ to the corresponding monochromatic luminosity $\nu L_\nu(\lambda)$ is given by $\log(\nu L_\nu/[L_\odot]) = -0.4 M_{\lambda, \rm AB} - \log (\lambda/[{\rm \mathring A}]) + 5.530$.} The contribution of type-2 AGNs at $z<1.5$ strongly increases from the \textit{g}- to the \textit{J}-band. Apart from the low-luminosity portion of the \textit{J}-band luminosity function, very likely affected by incompleteness, the agreement between the model and the data is remarkably good.


The comparisons between the global (stellar plus AGN components) luminosity functions yielded by the model and those observationally determined at several redshifts and wavelengths are shown in Figs.~\ref{fig:LFnu_15}--\ref{fig:LFnu_local_submm}. In Fig.~\ref{fig:zdistr} we compare model and observed redshift distributions at various wavelengths and flux density limits. The comparisons for all the other wavelengths for which estimates of the luminosity function are available can be found in the Web site  http://people.sissa.it/$\sim$zcai/galaxy\_agn/.

Note that a substantial fraction of sources have only photometric redshifts. For example, the fraction of photometric redshifts is 91\% for the VVDS-SWIRE survey with $S_{24\,\mu\rm m}>0.4\,$mJy \citep{Rodighiero2010}, 67.5\% for the GOODS-N and 36\% for the GOODS-S samples with $S_{24\,\mu\rm m}>0.08\,$mJy \citep{Rodighiero2010}. Only few sources at $z>2$ have spectroscopic redshifts \citep{Berta2011}. Note that photometric redshift errors tend to moderate the decline of the distributions at high-$z$. The effect is analogous to the Eddington bias on source counts: errors move more objects from the more populated lower $z$ bins to the less populated higher $z$ bins than in the other way. Thus the observed distributions may be overestimated at the highest redshifts. In addition, optical identifications are not always complete. On the whole, observational estimates of luminosity functions and of redshift distributions may be affected by systematic effects difficult to quantify and the true uncertainties may be larger than the nominal values.

%
%

\subsection{Model versus observed source counts and contributions to the CIB}

Model and observed source counts at wavelengths from $15\,\mu$m to 1.38\,mm are compared in Fig.~\ref{fig:nc_dnc}. At wavelengths $\ge 350\,\mu$m, where, in the present framework, proto-spheroidal galaxies are most important, the model provides a simple physical explanation of the steeply rising portion of the counts, that proved to be very hard to account for by other both physical \citep{Hayward2012,Niemi2012,Lacey2010} and phenomenological \citep[e.g.][]{Bethermin2012a,Gruppioni2011} models.

In our model the sudden steepening of the (sub-)mm counts is due to the appearance of proto-spheroidal galaxies that show up primarily at $z\gsim 1.5$, being mostly in passive evolution at lower redshifts. Their counts are extremely steep because, due to the strongly negative K-correction, the sub-mm flux densities corresponding to a given luminosity are only weakly dependent on the source redshift. Then, since the far-IR luminosity is roughly proportional to the halo mass, the counts reflect the high-$z$ luminosity function whose bright end reflects, to some extent, the exponential decline of halo mass function at high masses. This situation results in a very strong magnification bias due to gravitational lensing \citep{Blain1996,Perrotta2002,Perrotta2003,Negrello2007}. The counts of strongly lensed galaxies depend on the redshift distribution of the unlensed ones. Thus, the good agreement between the model and the observed counts of strongly lensed galaxies (see the $350\,\mu$m, $500\,\mu$m and $1380\,\mu$m panels of Fig.~\ref{fig:nc_dnc}) indicates that the model passes this test on the redshift distribution.

Low-$z$ ``warm'' and ``cold'' star-forming galaxy populations become increasingly important with decreasing wavelength. At $\lambda \ge 160\,\mu$m proto-spheroidal galaxies yield only a minor contribution to the counts. The AGN (mostly type-2) contribution implied by the model is always sub-dominant. We find a maximum contribution in the mid-IR. At $15\,\mu$m it is $\simeq 8\%$ up to 1 mJy and then rapidly increases up to $\simeq 20\%$ above 10 mJy while at $24\,\mu$m it is $\simeq 7$--$8\%$ up to 0.5 mJy and increases up to $\simeq 20\%$ above 2 mJy, in fair agreement with the observational estimates \citep{Treister2006,Teplitz2011}.

Another test on the redshift distribution is provided by the estimated counts in different redshift slices (Fig.~\ref{fig:nc_dnc_zbin}), although we caution that the true uncertainties may be larger than the nominal ones since the observational estimates are partly based on photometric redshifts and on stacking. The consistency between the model and the data is reasonably good.

Figure~\ref{fig:CIB} shows the contributions of the different populations to the cosmic infrared background (CIB). The model accounts for the full CIB intensity over the whole wavelength range. Only at $\lambda \le 10\,\mu$m other galaxy populations, such as passively evolving galaxies, become important. According to the model, for $\lambda \ge 350\,\mu$m the main contribution to the CIB comes from proto-spheroidal galaxies and the fraction contributed by these objects increases with increasing wavelengths. Below $\lambda = 350\,\mu$m lower $z$ ``warm'' galaxies take over, with ``cold'' galaxies adding a minor contribution. AGNs are always sub-dominant. The model gives a total (type-1 $+$ type-2 $+$ type-3) AGN contribution of 8.6\% at $16\,\mu$m and of 8.1\% at $24\,\mu$m.  For comparison, \citet{Teplitz2011} estimate a contribution of $\sim 15\%$ at $16\,\mu$m; \citet{Treister2006} and \citet{BallantynePapovich2007} find a contribution of $\sim 10\%$ at $24\,\mu$m. It must be noted that these observational estimates are endowed with substantial uncertainties: on one side they may be too low because strongly obscured AGNs may be missed, on the other side they may be too high because a significant fraction of the observed emission may come from the host galaxy.

\section{Clustering properties of dusty galaxies and power spectra of the cosmic infrared background}\label{sect:power_spectra}

An important test of our physical model for the evolution of dusty proto-spheroidal galaxies is provided by their clustering properties that are informative on their halo masses. A specific prediction of our model is that proto-spheroidal galaxies are the main contributors to the CIB at (sub-)mm wavelengths with ``warm'' starburst galaxies becoming increasingly important with decreasing wavelength. Since, in our model, proto-spheroidal galaxies are much more strongly clustered than starburst galaxies, the variation in the mixture with wavelength translates in quantitative predictions on the frequency dependence of the amplitude of the CIB power spectra and on the level of correlations among the maps at different frequencies.

We have updated the analysis by \citet{Xia2012} taking into account the new auto- and cross-frequency power spectra obtained by \citet{Viero2012} from {\it Herschel}/SPIRE measurements and the power spectrum at $100\,\mu$m derived by \citet{Penin2012}. The latter authors actually give also an estimate of the power spectrum at $160\,\mu$m. However the amplitude of the latter is anomalously large. As an example, for the wave-number $k_\theta = 0.03\,\hbox{arcmin}^{-1}$ we find that the amplitude normalized to the CIB intensity
\begin{equation}
{\delta I\over I} = {[2\pi k_\theta^2 P(k_\theta)]^{1/2}\over I_{\rm CIB}},
\end{equation}
[eq.~(13) of \citet{Viero2012}] is $\simeq 0.08$--0.09 at 100, 250, 350 and $500\,\mu$m but jumps to $\simeq 0.2$ at $160\,\mu$m.  Since such a jump over a small wavelength range looks odd we decided not to use the $160\,\mu$m power spectrum.

All the relevant details on the formalism used are given by \citet{Xia2012}. {Briefly, the power spectrum of the galaxy distribution is parameterized as the sum of the 1-halo term, that dominates on small scales and depends on the distribution of galaxies within the same
halo, and the 2-halo term, that dominates on large scales and is related to correlations among different halos. The Halo Occupation Distribution (HOD), which is a statistical description of how dark matter halos are populated with galaxies, is modeled using a central-satellite formalism \citep[see, e.g.,][]{Zheng2005}. This assumes that the first galaxy to be hosted by a halo lies at its
center, while any remaining galaxies are classified as satellites and are distributed in proportion to the halo mass profile. The mean halo occupation function of satellite galaxies is parameterized as: $\langle{N_{\rm sat}}\rangle \propto (M_{\rm vir}/M_{\rm sat})^{\alpha_{\rm sat}}$, where $M_{\rm vir}$ is the halo mass and the power-law index $\alpha_{\rm sat}$ is a free parameter. The key parameter in the 2-halo term is the minimum halo mass, $M_{\rm vir, min}$, that determines the amplitude of the effective bias function $b_{\rm eff}(z)$.}

In the \citet{Xia2012} paper the only free parameters are the minimum halo mass, $M_{\rm min, protosph}$, and the power-law index of the mean occupation function of satellites, $\alpha_{\rm sat, protosph}$, of proto-spheroidal galaxies. This is because the contribution of late-type galaxies to the power spectra at $\lambda \ge 250\,\mu$m is always subdominant and therefore the parameters characterizing their clustering properties were poorly constrained. This is no longer true if we add the $100\,\mu$m power spectrum, which, however, still provides only weak constraints on $\alpha_{\rm sat, late-type}$. We therefore fixed that parameter to $\alpha_{\rm sat, late-type}=1$. The fits to the {\it Herschel}/SPIRE power spectra determined by \citet{Viero2012} give $\log(M_{\rm min, protosph}/M_\odot) = 12.15 \pm 0.04$ and $\alpha_{\rm sat, protosph} = 1.55 \pm 0.05$ ($1\,\sigma$ errors), close to the values found by \citet{Xia2012}. The $100\,\mu$m data do not constrain these parameters further but yield $\log(M_{\rm min, late-type}/M_\odot) = 11.0 \pm 0.06$. The nominal errors on each parameter have been computed marginalizing on the other and correspond to $\Delta \chi^2=1$. We caution that the true uncertainties are likely substantially higher than the nominal values, both because the model relies on simplifying assumptions that may make it too rigid and because of possible systematics affecting the data. Our value of $M_{\rm min, protosph}$ implies an effective halo mass [eq.~(17) of \citet{Xia2012}] at $z\simeq 2$ of proto-spheroidal galaxies, making up most of the CIB, $M_{\rm eff} \simeq 4.5\times 10^{12}\,M_\odot$. This value is close to the estimated halo mass of the most effective star formers in the universe. \citet{Tacconi2008} estimated their mean comoving density at $z\sim 2$ to be $\sim 2\times 10^{-4}\,\hbox{Mpc}^{-3}$. For the standard $\Lambda$CDM cosmology this implies that they are hosted by dark matter halos of $\sim 3.5\times 10^{12}\,M_\odot$.

The best fit model power spectra are plotted in Fig.~\ref{fig:pk} where the 1- and 2-halo contributions of proto-spheroidal and late-type galaxies are also shown. The relative contribution of the latter galaxy population increases with decreasing wavelength and becomes dominant at $100\,\mu$m. This trend implies a decrease of the level of correlations among the maps with increasing separation in wavelength. As illustrated by Fig.~\ref{fig:norm_cross_spectra} the model is in very good agreement with the cross-wavelength correlations measured by \citet{Viero2012} and defined by [eq.~(14) of \citet{Viero2012}]
\begin{equation}
C_{A\times B} = {P^{A\times B}_{k_\theta} \over \sqrt{P^A_{k_\theta}\cdot P^B_{k_\theta}}}.
\end{equation}

\section{Summary and conclusions}\label{sect:conclusions}

Studies of galaxy properties as a function of morphological type \citep[e.g.][]{Bernardi2010} have highlighted a dichotomy between the luminosity-weighted ages of early- and late-type galaxies. The former are mostly older than 8 Gyr while most of S$b$ or later-type spirals are younger than $7$ Gyr, corresponding to a formation redshift $z\la 1$--1.5. Building on this datum we have worked out a model whereby the proto-spheroidal galaxies, in the process of forming the bulk of their stars, are the dominant population in the IR at $z\gsim 1.5$ while late-type galaxies dominate at lower redshifts. The model is  `hybrid' in the sense that it combines a physical, forward model for spheroidal galaxies and the early evolution of the associated AGNs with a phenomenological backward model for late-type galaxies and for the later AGN evolution.

To describe the cosmological evolution of proto-spheroidal galaxies and of the associated AGNs we adopted the physical model by \citet{Granato2004}, upgraded working out, for the first time, the epoch-dependent luminosity functions of sources as a whole (stellar plus AGN component), taking into account in a self-consistent way the variation with galactic age of the global SED. With only minor adjustments of the parameters the model accurately reproduces the observed luminosity functions at all redshifts ($z\gsim 1.5$) and IR wavelengths at which they have been determined. The model naturally accounts for the observed positive evolution of both galaxies and AGNs up to $z\simeq 2.5$ and for the negative evolution at higher redshifts. This is the result of the combination of two competing effects. On one side cooling and free-fall timescales shorten with increasing redshift because of the increase of the matter density and this yields higher star formation rates, i.e. higher galaxy luminosities at given halo mass. The higher gas densities are also responsible for a delay of the AGN switch-off time by feedback implying positive luminosity and density evolution of these objects. These effects are thwarted by the decrease in the comoving density of massive halos that prevails above $z\simeq 2.5$ causing a decline of the bolometric luminosity functions of both galaxies and AGNs.

The model also provides a simple physical explanation of the steeply rising portion of the (sub-)mm counts, that proved to be very hard to account for by other physical and phenomenological models. The sharp steepening is due to the sudden appearance of proto-spheroidal galaxies that do not have, in this spectral band, an evolutionary connection with nearby galaxies because their descendants are in passive evolution at $z\lsim 1.5$. Their (sub-)mm counts are extremely steep because, due to the strongly negative K-correction, the flux densities corresponding to a given luminosity are only weakly dependent on the source redshift. Then, since the far-IR luminosity is roughly proportional to the halo mass, the counts reflect, to some extent, the exponential decline of halo mass function at high masses.

The steepness of the counts imply a strong magnification bias due to gravitational lensing. The counts of strongly lensed sources depend on the redshift distribution that determines the distribution of lensing optical depths. In fact, this model was the only one that correctly predicted \citep{Negrello2007} the strongly lensed counts at $500\,\mu$m and the correct redshift distribution of bright ($S_{500\,\mu\rm m}\ge 100\,$mJy) sub-mm sources \citep{Negrello2010,Gonzalez-Nuevo2012}.

The epoch-dependent luminosity function of late-type galaxies has been modeled in terms of two populations, ``warm'' and ``cold'' galaxies with different SEDs and different evolution properties. Simple truncated power law models have been adopted for the evolution of these populations. ``Cold'' (normal) late-type galaxies evolve (weakly) only in luminosity, while ``warm'' (starburst) galaxies evolve both in luminosity and in density.

Below $z=1.5$ the far-IR emission of proto-spheroidal galaxies and the associated AGNs fade out rather rapidly. The AGNs, however, can be reactivated e.g. by interactions. This later phase of AGN emission has been described by a phenomenological model analogous to that used for late-type galaxies, distinguishing between type-1 and type-2 AGNs.

In this framework, there is a systematic variation with wavelength of the populations dominating the counts and the contributions to the extragalactic background intensity. Above $350\,\mu$m the main contributors to the CIB are proto-spheroidal galaxies. In this wavelength range late-type galaxies dominate the counts only at the brightest (where normal ``cold'' star-forming galaxies prevail) and at the faintest flux densities (where ``warm'' starburst galaxies outnumber the proto-spheroids). But these galaxies become increasingly important with decreasing wavelength. Proto-spheroids are always subdominant below $250\,\mu$m. This strong variation with wavelength in the composition of IR sources implies specific predictions for the auto- and cross-power spectra of the source distribution, that may help discriminating between different models. Essentially all the alternative models have all source populations present over the full relevant redshift range. This implies a high correlation between the CIB intensity fluctuations at different frequencies. On the contrary, the present model predicts a high (close to unity) cross-correlation  only at the longest wavelengths ($\ge 500\,\mu$m). At shorter wavelengths the cross correlation progressively weakens and we expect little cross-correlation between CIB fluctuations at, say, 100 and $500\,\mu$m. No observational determination is available for correlations among these wavelengths, but in the {\it Herschel}/SPIRE wavelength range, where cross correlations have been measured, the model results are in good agreement with observations.

According to our model, the AGN contribution to the CIB is always sub-dominant. It is maximal in the mid-IR where it reaches 8.6\% at $16\,\mu$m and 8.1\% at $24\,\mu$m. These contributions are close to, but somewhat lower than most observation-based estimates which however are complicated by the difficulty of separating the AGN emission from that of the host galaxy. The AGN contribution to the counts is also always subdominant. We find a maximum contribution in the mid-IR where the model gives AGN fractions in fair agreement with the observational estimates \citep{Treister2006,Teplitz2011}.

\begin{acknowledgements}
We are indebted to Matthieu B\'ethermin for several useful clarifications on flux calibration and color correction issues, to Roberto Assef for having sent his IR luminosity functions of AGNs in tabular form and to Aurelie P\'enin for having provided a tabulation of CIB power spectra at 100 and $160\,\mu$m and clarifications on error estimates. We also benefited from useful comments from an anonymous referee. Z.Y.C. acknowledges support from the joint PhD project between XMU and SISSA. A.L. thanks SISSA for warm hospitality. J.Q.X. is supported by the National Youth Thousand Talents Program and the grant No. Y25155E0U1 from IHEP. The work has been supported in part by ASI/INAF agreement n. I/072/09/0 and by INAF through the PRIN 2009 ``New light on the early Universe with sub-mm spectroscopy''.
\end{acknowledgements}

\appendix


\section{Self-regulated evolution of high-$z$ proto-spheroidal galaxies}\label{sect:appendix} %

The gas initially associated to a galactic halo of mass $M_{\rm vir}$, with a cosmological mass fraction $f_{\rm b} = M_{\rm gas}/M_{\rm vir}=0.165$ is heated to the virial temperature at the virialization redshift, $z_{\rm vir}$. Its subsequent evolution partitions it in three phases: a hot diffuse medium  with mass $M_{\mathrm{inf}}$ infalling and/or cooling toward the center; cold gas with mass $M_{\mathrm{cold}}$ condensing into stars; low-angular momentum gas with mass $M_{\mathrm{res}}$ stored in a reservoir around the central super-massive black hole, and eventually viscously accreting onto it. In addition, two condensed phases appear and grow, namely, stars with a total mass $M_{\star}$ and the black hole with mass $M_\bullet$. As mentioned in Section\,\ref{sect:overview} we restrict ourselves to the ranges $11.3 \lesssim \log(M_{\rm vir}/M_\odot) \lesssim 13.3$ and $z_{\rm vir} \gtrsim 1.5$.


The evolution of the three gas phases is governed by the following equations:
\begin{eqnarray}
\dot{M}_{\rm inf} &=& - \dot{M}_{\rm cond} - \dot{M}_{\rm inf}^{\rm QSO},\nonumber \\
\dot{M}_{\rm cold} &=& \dot{M}_{\rm cond} - [1-{\cal{R}}(t)]\dot{M}_{\star} - \dot{M}_{\rm cold}^{\rm SN} - \dot{M}_{\rm cold}^{\rm QSO},\\
\dot{M}_{\rm res} &=& \dot{M}_{\rm inflow} - \dot{M}_{\rm BH},\nonumber
\end{eqnarray}
that link the mass infall rate, $\dot{M}_{\rm inf}$, the variation of the cold gas mass, $\dot{M}_{\rm cold}$, and the variation of the reservoir mass, $\dot{M}_{\rm res}$,  to the condensation rate of the cold gas, $\dot{M}_{\rm cond}$, to the star formation rate $\dot{M}_\star$, to the cold gas removal by supernova and AGN feedback, $\dot{M}_{\rm cold}^{\rm SN}$ and $\dot{M}_{\rm cold}^{\rm QSO}$ respectively, to the fraction of gas restituted to the cold component by the evolved stars, $\mathcal{R}(t)$, to the inflow rate of cold gas into the reservoir around the central super-massive black hole, $\dot{M}_{\rm inflow}$, and to the back hole accretion rate, $\dot{M}_{\rm BH}$.

{The hot gas cools and flows toward the central region at a rate}
%
\begin{equation}
	\dot{M}_{\rm cond} \simeq \frac{M_{\rm inf}}{t_{\rm cond}},
\end{equation}
with $M_{\rm inf}^{0}=f_{\rm b} M_{\rm vir}$ and
\begin{equation}
	t_{\rm cond} \simeq 8\times10^8 \Big(\frac{1+z}{4}\Big)^{-1.5} \Big(\frac{M_{\rm vir}}{10^{12}M_{\odot}}\Big)^{0.2}\ {\rm yr},
\end{equation}
where the coefficient is 10\% smaller than the value used by \citet{Fan2010}. {Note that the cooling and inflowing gas we are dealing with is the one already present within the halo at virialization. In this respect it is useful to keep in mind that the virial radius of halo ($R_{\rm vir}\simeq 220 (M_{\rm vir}/10^{13}\,M_\odot)^{1/3}[3/(1+z_{\rm vir})]\,$kpc) is more than 30 times larger than the size of the luminous galaxy, and that only a minor fraction of the gas within the halo condenses into stars. Indeed, we need strong feedback processes, capable of removing most of the halo gas, to avoid an over-production of stars. This implies that any gas infalling from outside the halo must also be swept out by feedback; it could however become important for the formation of a disc-like structure surrounding the preformed spheroid once it enters the passive evolution phase, with little feedback \citep{Cook2009}. As mentioned in Sect.~\ref{sect:overview}, the additional material (stars, gas, dark matter) infalling after the fast collapse phase that creates the potential well, i.e. during the slow-accretion phase, mostly produces a growth of the halo outskirts, and has little effect on the inner part where the visible galaxy resides. }

The star formation rate is given by
\begin{equation}
	\dot{M}_\star \simeq \frac{M_{\rm cold}}{t_\star},
\end{equation}
where the star formation timescale is $t_\star \simeq t_{\rm cond}/s$ with $s \simeq 5$. For a \citet{Chabrier2003} IMF of the form $\phi(m)=m^{-x}$ with $x=1.4$ for $0.1 \leqslant m \leqslant 1 M_\odot$ and $x=2.35$ for $m > 1 M_\odot$ we find  ${\cal R} \simeq 0.54$ under the instantaneous recycling approximation.

The gas mass loss due to the supernova feedback is
\begin{equation}
	\dot{M}_{\rm cold}^{\rm SN} = \beta_{\rm SN} \dot{M}_\star,
\end{equation}
with
\begin{eqnarray}\label{eq:SN}
	\beta_{\rm SN} = &\mathlarger{\frac{N_{\rm SN} \epsilon_{\rm SN} E_{\rm SN}}{E_{\rm bind}}} \simeq 0.6 \mathlarger{\left(\frac{N_{\rm SN}}{8\times10^{-3}/M_\odot}\right)} \mathlarger{\left(\frac{\epsilon_{\rm SN}}{0.05}\right)} \nonumber\\
	&\times \mathlarger{\left(\frac{E_{\rm SN}}{10^{51}\ {\rm erg}}\right)} \mathlarger{\left(\frac{M_{\rm vir}}{10^{12}M_\odot}\right)^{-2/3}} \mathlarger{\left(\frac{1+z}{4}\right)^{-1}}.
\end{eqnarray}
We adopt the following values: number of SNe per unit solar mass of condensed stars $N_{\rm SN} \simeq 1.4\times10^{-2}/M_\odot$; fraction of the released energy used to heat the gas $\epsilon_{\rm SN} = 0.05$;  kinetic energy released per SN $E_{\rm SN} \simeq 10^{51}\ {\rm ergs}$; halo binding energy $E_{\rm bind} \simeq 3.2\times10^{14} (M_{\rm vir}/10^{12}M_\odot)^{2/3} ([(1+z)/4]\ \rm cm^2\ s^{-2}$ \citep{MoMao2004}.

The infrared luminosity (8--$1000\,\mu$m) associated to dust enshrouded star formation is
\begin{equation}\label{eq:SFR-IR}
	L_{\star,\rm IR}(t) = k_{\star,\rm IR} \times10^{43} \Big(\frac{\dot{M}_\star}{M_\odot\ {\rm yr}^{-1}}\Big)\ {\rm erg\ s}^{-1},
\end{equation}
where the coefficient $k_{\star,\rm IR}$ depends on the SED. We adopt $k_{\star,\rm IR} \sim 3$ \citep{Lapi2011,Kennicutt1998}.

The cold gas inflow rate into the reservoir around the super-massive black hole, driven by radiation drag, is given by
\begin{equation}\label{eq:RD}
	\dot{M}_{\rm inflow} \simeq \frac{L_\star}{c^2}(1-e^{-\tau_{\rm RD}}) \simeq \alpha_{\rm RD} \times 10^{-3} \dot{M}_\star (1-e^{-\tau_{\rm RD}}),
\end{equation}
with
\begin{equation}\label{eq:optical_depth}
	\tau_{\rm RD}(t) = \tau_{\rm RD}^0 \Big(\frac{Z_{\rm cold}(t)}{Z_\odot}\Big) \Big(\frac{M_{\rm cold}}{10^{12}M_\odot}\Big) \Big(\frac{M_{\rm vir}}{10^{13}M_\odot}\Big)^{-2/3}.
\end{equation}
For the strength of the radiation drag we adopt $\alpha_{\rm RD} = 2.5$ and set $\tau^0_{\rm RD} = 3.0$. The model also follows the evolution of the cold gas metallicity, $Z_{\rm cold}(t)$. An approximate solution of the equations governing the chemical evolution is (Lapi et al., in preparation)
\begin{eqnarray}\label{eq:Zcold}
	Z_{\rm cold}  (t) = Z^0_{\rm inf} + \frac{s}{s \gamma - 1}{\cal E}_Z(t) - \frac{s t / t_{\rm cond}}{e^{(s \gamma -1)t/t_{\rm cond}} -1} \cdot \nonumber\\
	 \cdot  \Big\{ {\cal E}_Z(t) + B_Z \sum^{\infty}_{i=2} \frac{1}{i \cdot i!} \Big[ (s \gamma -1) \frac{\min(t, t_Z)}{t_{\rm cond}} \Big]^{i-1} \Big\},
\end{eqnarray}
where $\gamma = 1 - {\cal R} - \beta_{\rm SN}$, the metallicity of the primordial infalling gas is $Z^0_{\rm inf} = 10^{-5}$, and the mass fraction of newly formed metals ejected from stars, ${\cal E}_Z(t)$ is given by
\begin{equation}\label{eq:metal_yield}
 \simeq A_Z + B_Z \ln \Big[ \frac{\min(t, t_{\rm saturation})}{t_Z} \Big]
\end{equation}
with $A_Z = 0.03$, $B_Z = 0.02$, $t_Z = 20$ Myr, and $t_{\rm saturation} = 40$ Myr for the Chabrier's IMF ($Z_\odot \simeq 0.02$). {
Equation~(\ref{eq:metal_yield}) accounts for the fact that, soon after the onset of star formation, the metal yield, mainly contributed by stars with large masses ($\ge 20\,M_\odot$) and short lifetimes ($t_Z\le 20\,$Myr), is  a relatively large fraction of the initial stellar mass ($f_{\rm metal} \ge 0.12$) while, as the star formation proceeds, it progressively lowers to $f_{\rm metal} \sim 0.06$ as the main contribution shifts to stars with intermediate masses $\sim 9-20\ M_\odot$ and lifetimes $t_Z \sim 20-40\,$Myr, and finally
saturates to values $f_{\rm metal} \sim 0.013$ as stars with masses $\le 9\ M_\odot$ and long lifetimes ($t_{\rm saturation} \ge 40$ Myr) take over \citep{Bressan1998}. The two parameters $A_Z$ and $B_Z$ depends mainly on the IMF.}

The accretion rate into the central black hole obeys the equation
\begin{equation}
	\dot{M}_{\rm BH} = \min(\dot{M}_{\rm BH}^{\rm visc}, \lambda_{\rm Edd} \dot{M}_{\rm Edd}),
\end{equation}
where $\dot{M}_{\rm BH}^{\rm visc}$ is the accretion rate allowed by the viscous dissipation of the angular momentum of the gas in the reservoir
\begin{eqnarray}
	\dot{M}_{\rm BH}^{\rm visc} = \frac{M_{\rm res}}{\tau_{\rm visc}} &= \kappa_{\rm accr}\ 5\times10^3 \mathlarger{\left(\frac{V_{\rm vir}}{500\ {\rm km\ s}^{-1}}\right)^3} \nonumber \\
	& \times \mathlarger{\left(\frac{M_{\rm res}}{M_{\bullet}}\right)^{3/2}} \mathlarger{\left(1+\frac{M_{\bullet}}{M_{\rm res}}\right)^{1/2}},
\end{eqnarray}
with $\kappa_{\rm accr} \simeq 10^{-2}$ and $V_{\rm vir}^2 = G M_{\rm vir}^{2/3} [4\pi \Delta_{\rm vir}(z) \bar\rho_{\rm m}(z)/3]^{1/3}$, $\Delta_{\rm vir}$ being the overdensity of a virialized halo at redshift $z_{\rm vir}$ within its virial radius $r_{\rm vir}$. $\dot{M}_{\rm Edd} \equiv M_\bullet/\epsilon\, t_{\rm Edd}$ is the accretion rate corresponding to the Eddington luminosity given the mass to light conversion efficiency $\epsilon$ (we set $\epsilon=0.1$ so that the Salpeter time $\epsilon\, t_{\rm Edd}=4.5 \times 10^7$ yr) and $\lambda_{\rm Edd}(z)$ is the Eddington ratio that we assume to slightly increase with redshift for $z \gtrsim 1.5$
\begin{equation}\label{eq:Edd_ratio}
\lambda_{\rm Edd}(z) \simeq 0.1 (z-1.5)^2 + 1.0
\end{equation}
up to a maximum value $\lambda_{\rm Edd, max}=4$. The growth rate of the black hole mass is
\begin{equation}
	\dot{M}_{\bullet}(t) = (1-\epsilon) \dot{M}_{\rm BH}
\end{equation}
starting from a seed mass $M_{\bullet}^{\rm seed}=10^2\,M_\odot$. The bolometric AGN luminosity is
\begin{equation}\label{eq:Lbh}
	L_{\bullet} = \epsilon \dot M_{\rm BH} c^2 = 5.67\times10^{45} \Big(\frac{\epsilon}{0.1}\Big) \Big(\frac{\dot M_{\rm BH}}{M_\odot\ \rm yr^{-1}}\Big)\ {\rm erg\ s^{-1}}.
\end{equation}
A minor fraction of it couples with the interstellar medium of the host galaxy giving rise to an outflow at a rate %
\begin{equation}
	\dot{M}_{\rm inf,cold}^{\rm QSO} = \dot{M}_{\rm wind} \frac{M_{\rm inf,cold}}{M_{\rm inf}+M_{\rm cold}},
\end{equation}
with
\begin{equation}
	\dot{M}_{\rm wind} = \frac{L_{\rm QSO}^{\rm ISM}}{E_{\rm bind}},
\end{equation}
and
\begin{equation}\label{eq:L_QSO}
	L_{\rm QSO}^{\rm ISM} \simeq 2\times10^{44}\epsilon_{\rm QSO}\Big(\frac{\dot M_{\rm BH}}{M_\odot\ {\rm yr}^{-1}}\Big)^{3/2}\ {\rm erg\ s}^{-1}.
\end{equation}
$L_{\rm QSO}^{\rm ISM}$ is the mechanical AGN luminosity, used to unbind the gas. The coefficient quantifying the strength of the QSO feedback is chosen to be $\epsilon_{\rm QSO}=3$. The ratio of the mechanical to the total AGN luminosity
\begin{equation}
L_{\rm QSO}^{\rm ISM}/L_{\bullet} \simeq 3.5\times10^{-3}\ \frac{\epsilon_{\rm QSO}}{\epsilon} \left(\frac{\dot M_{\rm BH}}{M_\odot\ {\rm yr}^{-1}}\right)^{1/2},
\end{equation}
is constrained to be in the range 0.006--0.15.

Examples of the resulting evolution with galactic age of properties of the stellar and of the AGN component are shown in Fig.~\ref{fig:sph_evol_mlz3} for three values of the virial mass and $z_{\rm vir}=3$.

{As mentioned in Sect.~\ref{sect:parameters}, to improve the fits of the data we have modified, by trial and error, the values of some model parameters used in previous papers, still within their plausible ranges (see Table~\ref{tab:parameters_proto}). The impact of these parameters on the derived luminosity functions can be more easily understood with reference to the time lag between the halo virialization and the peak in black hole accretion rate, $\Delta t_{\rm peak}$ \citep{Lapi2006}. The duration of star formation is $\Delta t_{\rm SF} \lesssim \Delta t_{\rm peak}$ (see Fig.~\ref{fig:sph_evol_mlz3}) due to the drastic effect of QSO feedback in massive halos which dominate the bright end of the luminosity functions. Note that longer $\Delta t_{\rm peak}$ (or $\Delta t_{\rm SF}$) imply higher bright tails of the luminosity functions. The final black hole mass increases with increasing the coefficient, $\tau^0_{\rm RD}$, of the optical depth of gas clouds [eq.~(\ref{eq:optical_depth})] because it implies a higher efficiency of the radiation drag driving the gas into the reservoir. There is a degeneracy, to some extent, between $\tau^0_{\rm RD}$ and the gas metallicity $Z_{\rm cold}$, implying that $\tau^0_{\rm RD}$ cannot be tightly constrained \citep[see][]{Granato2004}. The value of $\Delta t_{\rm peak}$ grows substantially in response to a small increase of the radiative efficiency $\epsilon$ that yields a slower growth of the black hole mass and a weaker QSO feedback. Higher values of the Eddington ratio,  $\lambda_{\rm Edd}$, result in lower values of both $\Delta t_{\rm peak}$ and of the final black hole mass. A rise of $\lambda_{\rm Edd}$ at  high-$z$ is required to account for the observed space density of very luminous QSOs \citep[see the high-$z$ data in  Fig.~\ref{fig:LFnu_g} and Fig.~\ref{fig:LFnu_J};][]{Lapi2006}. A higher QSO feedback efficiency (higher $\epsilon_{\rm QSO}$) shortens the duration of star formation, $\Delta t_{\rm SF}$, but has a minor effect on $\Delta t_{\rm peak}$ and on the final black hole mass. Finally, the coefficient relating the SFR to the IR luminosity, $k_{\star, \rm IR}$, varies with age mix of stellar populations, chemical composition and IMF. Increasing it we shift the luminosity functions towards higher luminosities. }

\clearpage

\begin{figure} 
\begin{center}
	\includegraphics[width=\columnwidth]{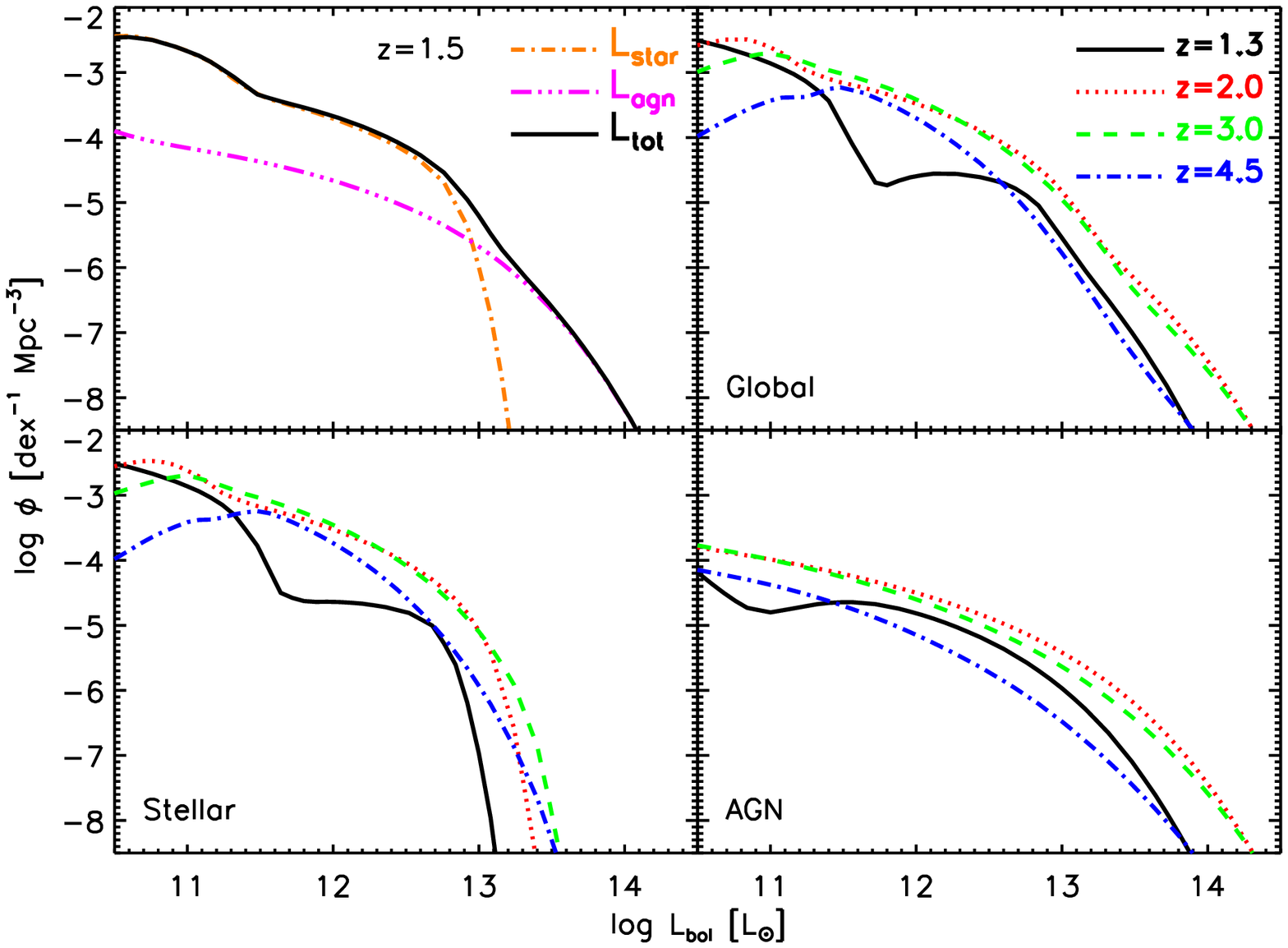}
	\caption{Bolometric luminosity functions of proto-spheroidal galaxies. The upper left panel shows the luminosity functions at $z=1.5$ of the stellar (dot-dashed orange line) and of the AGN component luminosity (triple-dot-dashed magenta line), as well as the global luminosity function (solid black line). Note that, as discussed in Section\,\protect\ref{sect:protoLF}, the latter is {\it not} the sum of the two components although, in this case, is very close to it. The upper right panel illustrates the evolution of the global luminosity function from $z=1.3$ to $z=4.5$, while the lower panels show the evolution of each component separately. The decline at low luminosities is an artifact due to the adopted lower limit to the proto-spheroid halo masses. The figure highlights the different shapes of the stellar and AGN bolometric luminosity function, with the latter having a more extended high luminosity tail, while the former sinks down exponentially above $\sim 10^{13}\,L_\odot$. The evolutionary behaviour of the two components is qualitatively similar and cannot be described as simple luminosity or density evolution; down-sizing effects are visible in both cases. On the other hand there are also clear differences.}\label{fig:sph_LFbol}
\end{center}
\end{figure}

\clearpage

\begin{figure} 
\begin{center}
	\includegraphics[width=\columnwidth]{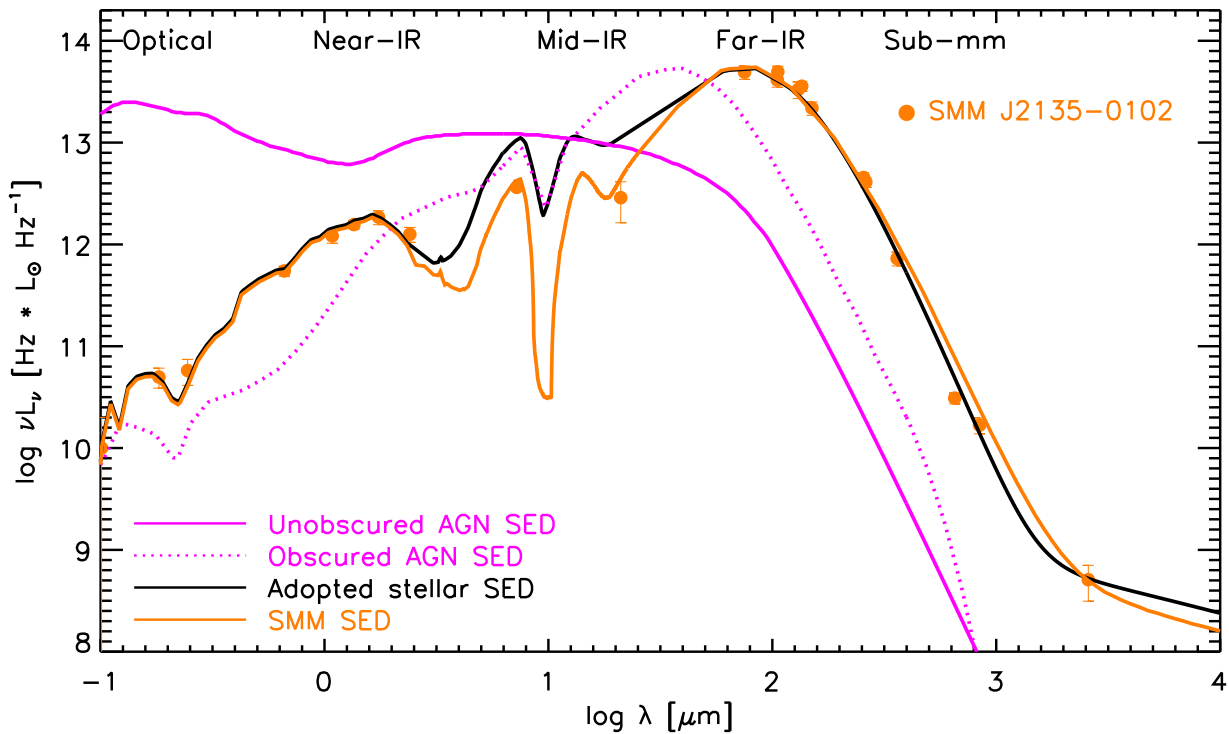}
\caption{SEDs of stellar and AGN components of proto-spheroidal galaxies. The solid black line shows the adopted SED for the stellar component, obtained modifying that of the $z\simeq 2.3$ galaxy SMM~J2135-0102, also shown for comparison [solid orange line; the photometric data data are from \citet{Swinbank2010} and \citet{Ivison2010}]. The dotted magenta line represents the SED adopted for the dust obscured phase of the AGN evolution and is taken from the AGN SED library by \citet{GranatoDanese1994}. For unobscured AGNs we have adopted the mean QSO SED of Richards et al. (2006; solid magenta line). The original SMM~J2135-0102 SED and the 2 AGN SEDs are normalized to $\log (L_{\rm IR}/L_\odot) = 13.85$ while the modified SED is normalized to $\log (L_{\rm IR}/L_\odot) = 13.92$ to facilitate the comparison with the original SED. Except in the rare cases in which the AGN bolometric luminosity is much larger than that of the starburst, the AGN contribution is small at (sub-)mm wavelengths, while it is important and may be dominant, in the mid-IR.}\label{fig:seds_sph}
\end{center}
\end{figure}

\clearpage

\begin{figure} 
\begin{center}
	\includegraphics[width=\columnwidth]{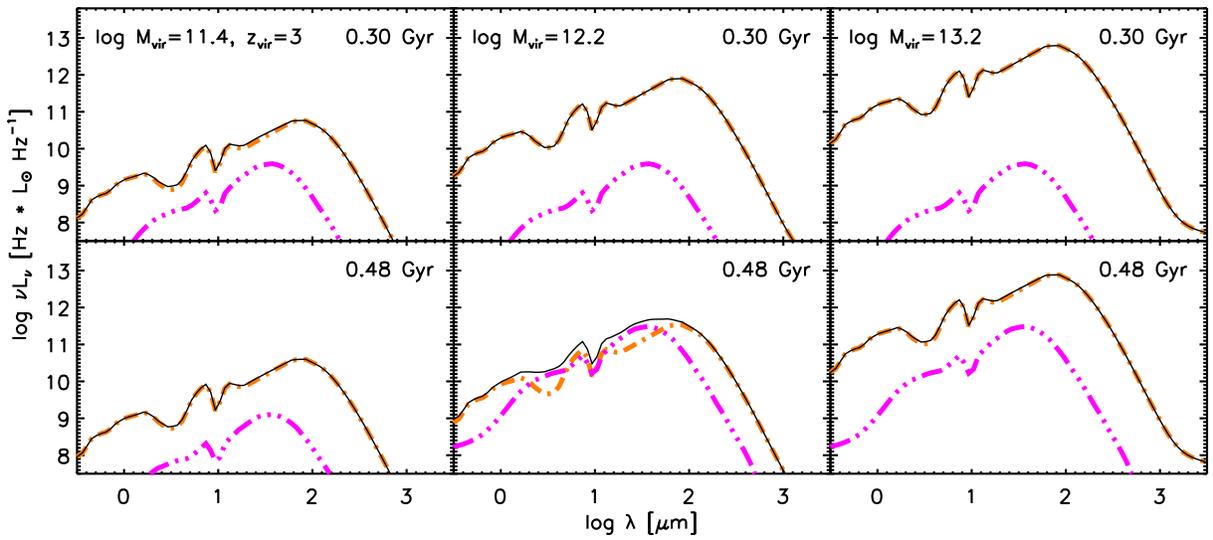}
	\caption{Global SEDs (solid black lines) for two galactic ages (0.3 and 0.48 Gyr) and three host halo masses ($\log(M_H/M_\odot)=11.4$, 12.2 and 13.2, from left to right), virialized at $z_{\rm vir} = 3$. The dot-dashed  orange line (overlaid by the solid black line in some panels) and the triple-dot-dashed magenta line show the stellar and the AGN component, respectively. The shorter evolution timescale of the AGNs is clearly visible. The effect of feedback as a function of halo mass on the SFR is very different from that on accretion onto the super-massive black-hole (see text).}\label{fig:seds_sph_tM}
\end{center}
\end{figure}

\clearpage

\begin{figure} 
\begin{center}
	\includegraphics[width=\columnwidth]{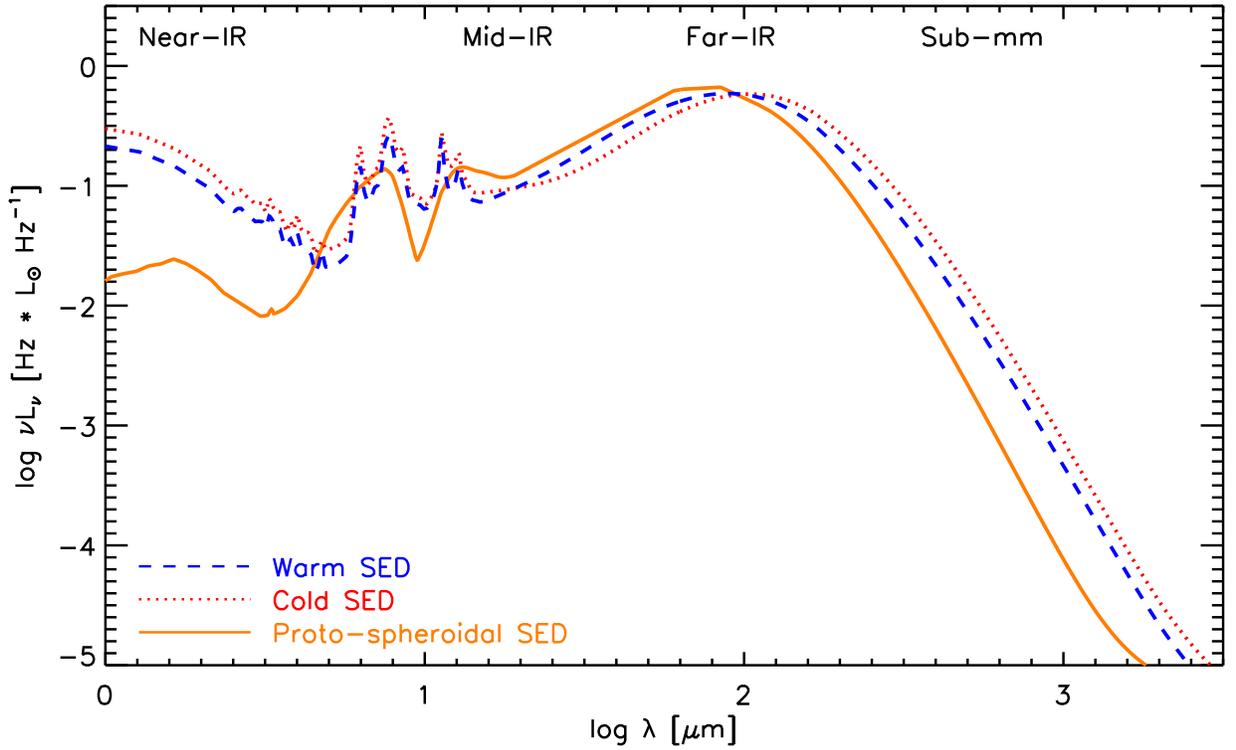}
	\caption{Adopted SEDs for the ``warm'' (dashed blue line) and ``cold'' (dotted red line) low-$z$ star-forming galaxies. They were generated combining SEDs of \citet{DaleHelou2002} and \citet{Smith2012}, as described in the text. The solid orange line shows, for comparison, the SED of proto-spheroidal galaxies.  The 3 SEDs are normalized to the same total IR luminosity $\log(L_{\rm IR}/L_\odot) = 1$.}\label{fig:seds_sfg}
\end{center}
\end{figure}

\clearpage

\begin{figure} 
\begin{center}
	\includegraphics[width=\columnwidth]{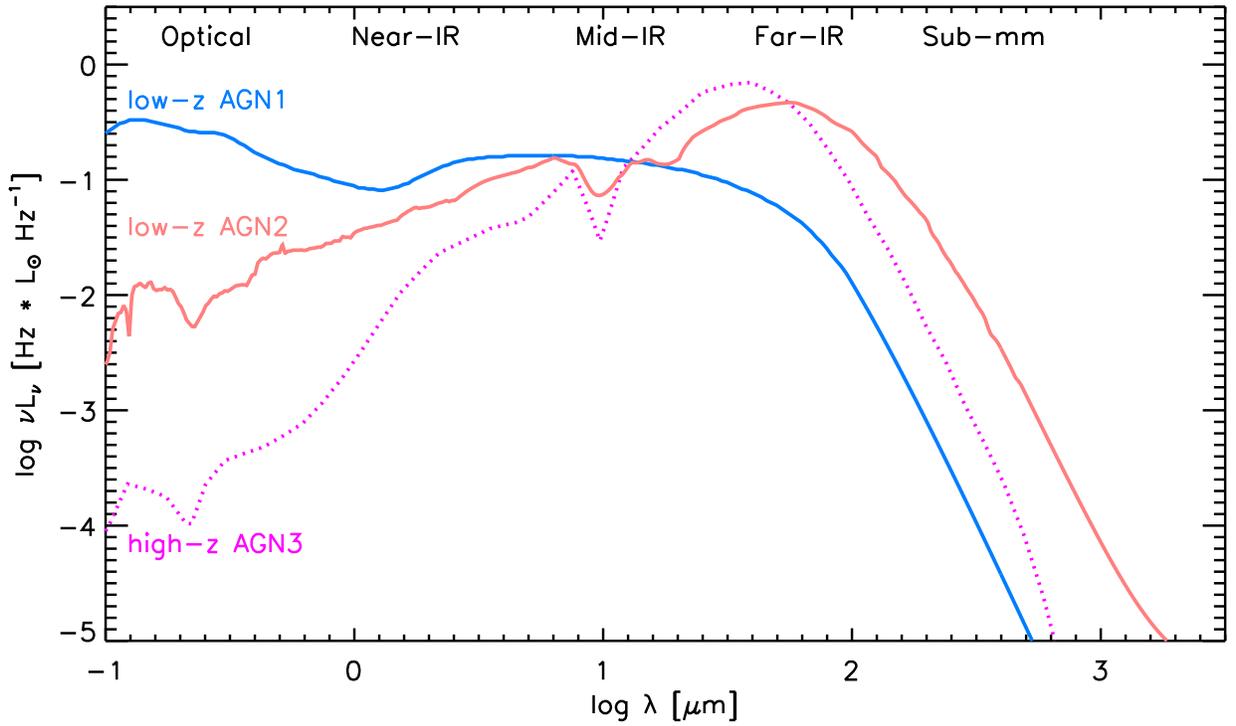}
	\caption{SEDs of low-$z$ type-1 AGNs (solid light-blue line) and type-2 AGNs (solid pink line). The dotted magenta line shows, for comparison, the adopted SED of AGNs associated to dusty proto-spheroidal galaxies (type-3 AGNs). The SEDs are normalized to the same, arbitrary, bolometric luminosity. }\label{fig:seds_agn}
\end{center}
\end{figure}

\clearpage

\begin{figure*} 
\begin{center}
	\includegraphics[width=\textwidth]{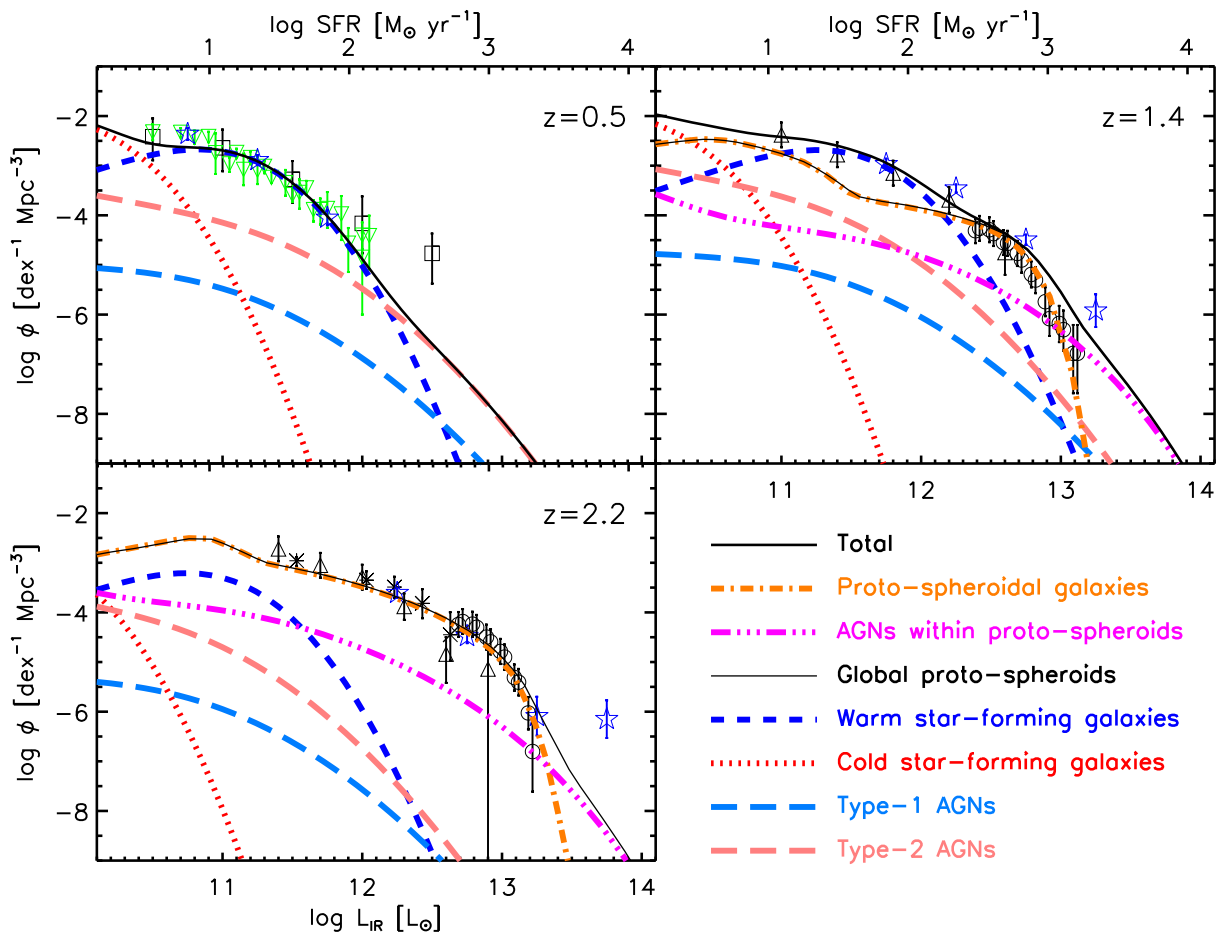}
	\caption{Comparison between model and observational determinations of the IR (8--$1000\,\mu$m) luminosity functions at several redshifts. At $z>1.0$ we have contributions from proto-spheroidal galaxies (dot-dashed orange lines) and from the associated AGNs (both obscured and unobscured; triple-dot-dashed magenta lines). The thin solid black lines (that are generally superimposed to the dot-dashed orange lines) are the combination of the two components. These contributions fade at lower redshifts and essentially disappear at $z<1$. At $z\le 1.5$ the dominant contributions come from ``warm'' (short-dashed blue lines) and ``cold'' (dotted red lines) star forming galaxies. Type-2 AGNs (long-dashed pink lines) or type-3 AGNs associated to dusty proto-spheroids (triple-dot-dashed magenta lines) dominate at the highest IR luminosities while type-1 AGNs (long-dashed light-blue lines) are always sub-dominant (in the IR). The thick solid black lines show the sum of all contributions. The upper horizontal scale gives an estimate of the SFRs corresponding to IR luminosities. These estimates are only indicative (see Sect.~\ref{sect:results_lf_rd}). Data points are from \citet[black open squares]{LeFloch2005}, \citet[black stars]{Caputi2007}, \citet[green downward triangles]{Magnelli2009}, \citet[blue open asterisks]{Rodighiero2010}, \citet[black triangles]{Magnelli2011}, and \citet[black open circles]{Lapi2011}. 
}\label{fig:LF_IR}
  \end{center}
\end{figure*}




\clearpage


\begin{figure*} 
\begin{center}
	\includegraphics[width=\textwidth]{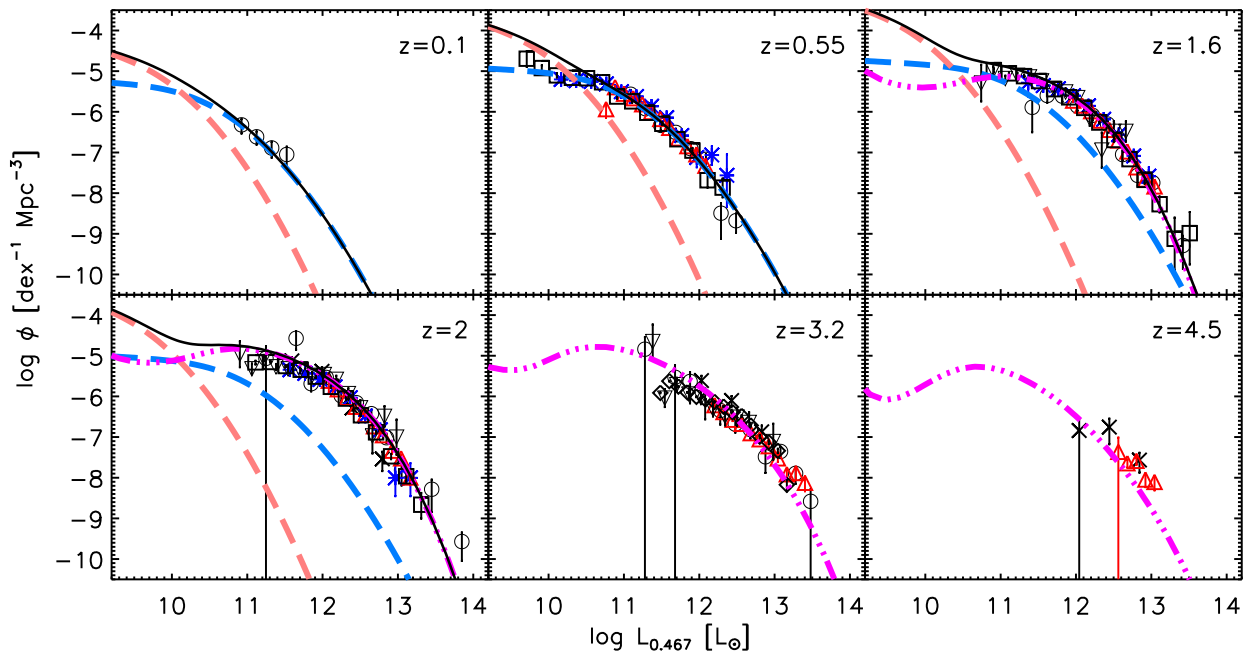}
	\caption{Comparison between model and observed {\textit g}-band ($0.467\,\mu$m) AGN luminosity function at several redshifts. As in Fig.~\protect\ref{fig:LF_IR} the long-dashed light-blue and pink lines refer to type-1 and type-2 AGNs, respectively, while the triple-dot-dashed magenta lines refer to AGNs associated with proto-spheroidal galaxies. At $z<2$ the solid black line shows the sum of all the contributions. At higher $z$ only proto-spheroids are considered. Data points are from \citet[black open circles]{HartwickSchade1990}, \citet[black crosses]{Warren1994}, \citet[blue stars]{Croom2004}, \citet[red triangles]{Richards2006}, \citet[black open squares]{Croom2009}, \citet[black downward triangles]{PalanqueDelabrouille2012}, and \citet[black diamonds]{Ross2012}. The data by \citet{HartwickSchade1990}, given in terms of $M_B$ in the Vega system, were converted to $M_g$ adopting the $B-g \simeq 0.14$ colour estimated by \citet{Fukugita1996} and were further corrected for the the different cosmology. The UV magnitudes of \citet{Warren1994} were first converted to $B$ magnitudes ($M_B=M_{C,1216\,\mathring {\rm A}}+1.39\alpha_\nu+0.09$, with $\alpha_\nu=-0.5$) following \citet{Pei1995} and then to $M_g$ as before. The data by \citet{Ross2012} were converted from $M_{i}(z=2)$ to $M_g$ following \citet{Richards2006} with spectral index $\alpha_\nu=-0.5$. The correction for the different cosmology was also applied. Finally, the conversion from $M_g$ to $\nu L_\nu(0.467\ \mu$m) is given in Section \ref{sect:results_lf_rd}.}\label{fig:LFnu_g}
\end{center}
\end{figure*}

\clearpage

\begin{figure*} 
\begin{center}
	\includegraphics[width=\textwidth]{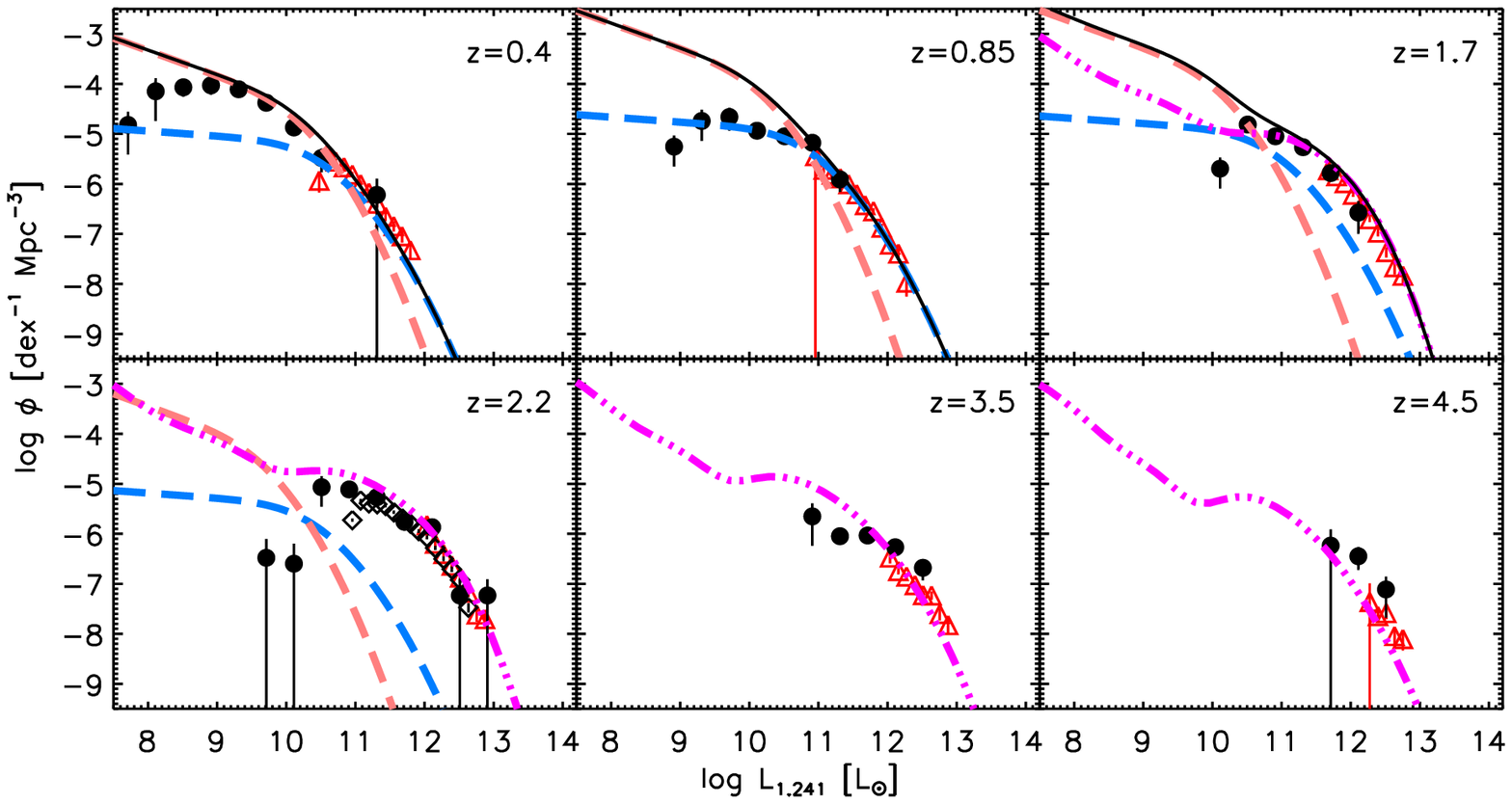}
	\caption{Comparison between model and observed {\textit J}-band ($1.24\,\mu$m) AGN luminosity function at several redshifts. The lines have the same meaning as in Fig.~\protect\ref{fig:LFnu_g}. There are clear signs of substantial incompleteness at the lowest luminosities. Data are from \citet[red triangles]{Richards2006}, \citet[black filled circles]{Assef2011}, and \citet[black diamonds]{Ross2012}. The data by \citet{Assef2011}, given in terms of $M_J$ in the Vega system, were converted to $\nu L_\nu (1.241\ \mu$m) using the relation ($L_\nu(1.241\ \mu{\rm m}) = 1623 \times 10^{-0.4 M_J}$ Jy) by \citet{Rieke2008}. The $i$-band data by \citet{Richards2006} and \citet{Ross2012} were converted to $M_{J,\rm AB}$ assuming spectral index $\alpha_\nu=-0.5$. }\label{fig:LFnu_J}
\end{center}
\end{figure*}





\clearpage

\begin{figure*} 
\begin{center}
	\includegraphics[width=\textwidth]{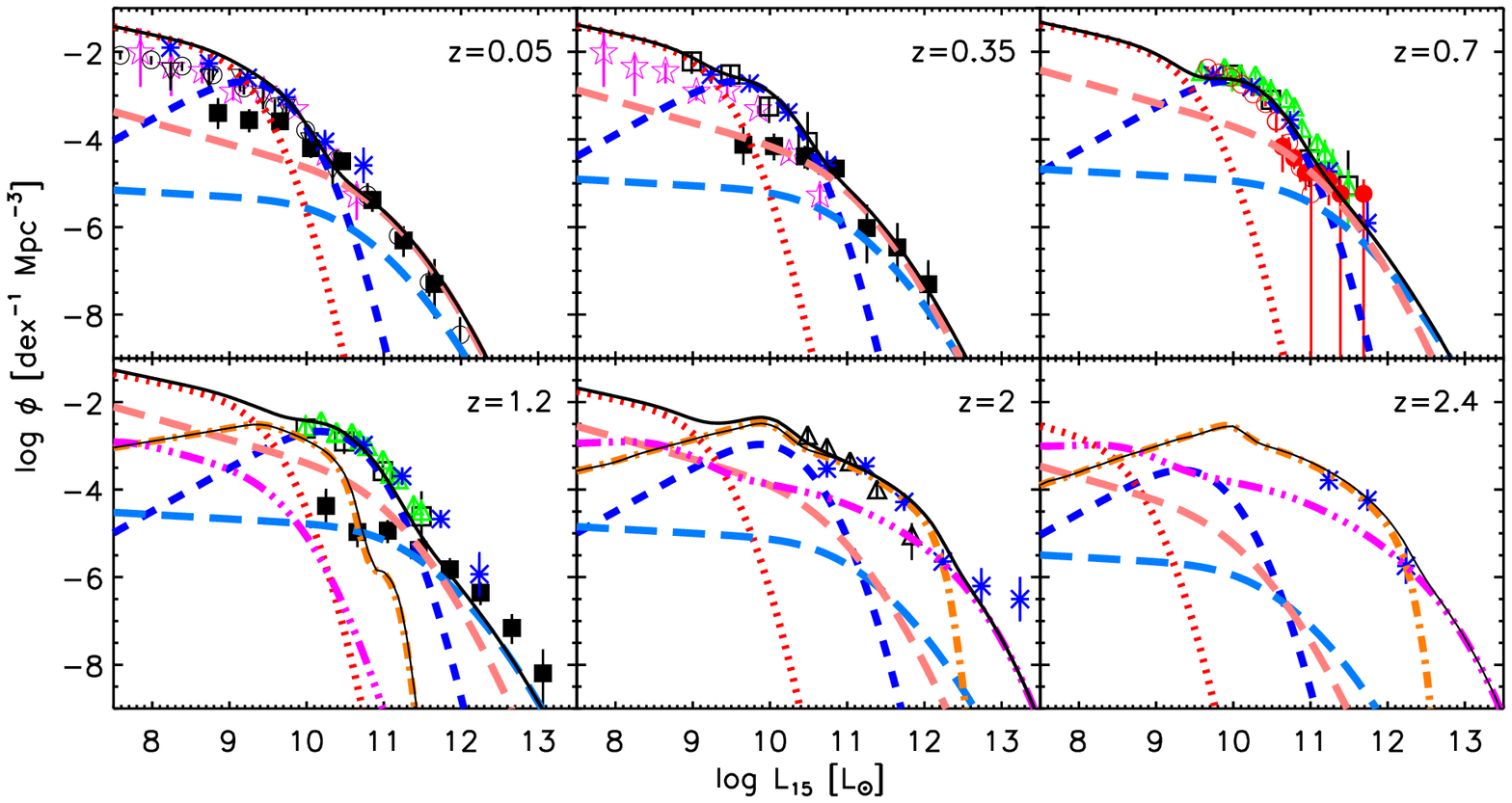}
	\caption{Comparison between model and observed $15\,\mu$m global (galaxies plus AGNs) luminosity function at several redshifts. Data are from \citet[magenta open asterisks]{Pozzi2004}, \citet[black open squares]{LeFloch2005}, \citet[black filled squares]{Matute2006}, \citet[black open circles]{Mazzei2007}, \citet[green triangles]{Magnelli2009}, \citet[blue stars]{Rodighiero2010}, \citet[red open circles for star-formation and red filled circles for AGNs]{Fu2010}, \citet[black downward triangles]{Wu2011}, and \citet[black triangles]{Magnelli2011}. The black filled squares in the panels at $z=0.05$ and 0.35 show observational estimates of the luminosity function of type-2 AGNs only while the red filled circles at $z=0.7$ refer to AGN of both types and at $z=1.2$ refer to type-1 only. The lines have the same definition as in Fig.~\ref{fig:LF_IR}.  }\label{fig:LFnu_15}
\end{center}
\end{figure*}







\clearpage

\begin{figure*} 
\begin{center}
	\includegraphics[width=\textwidth]{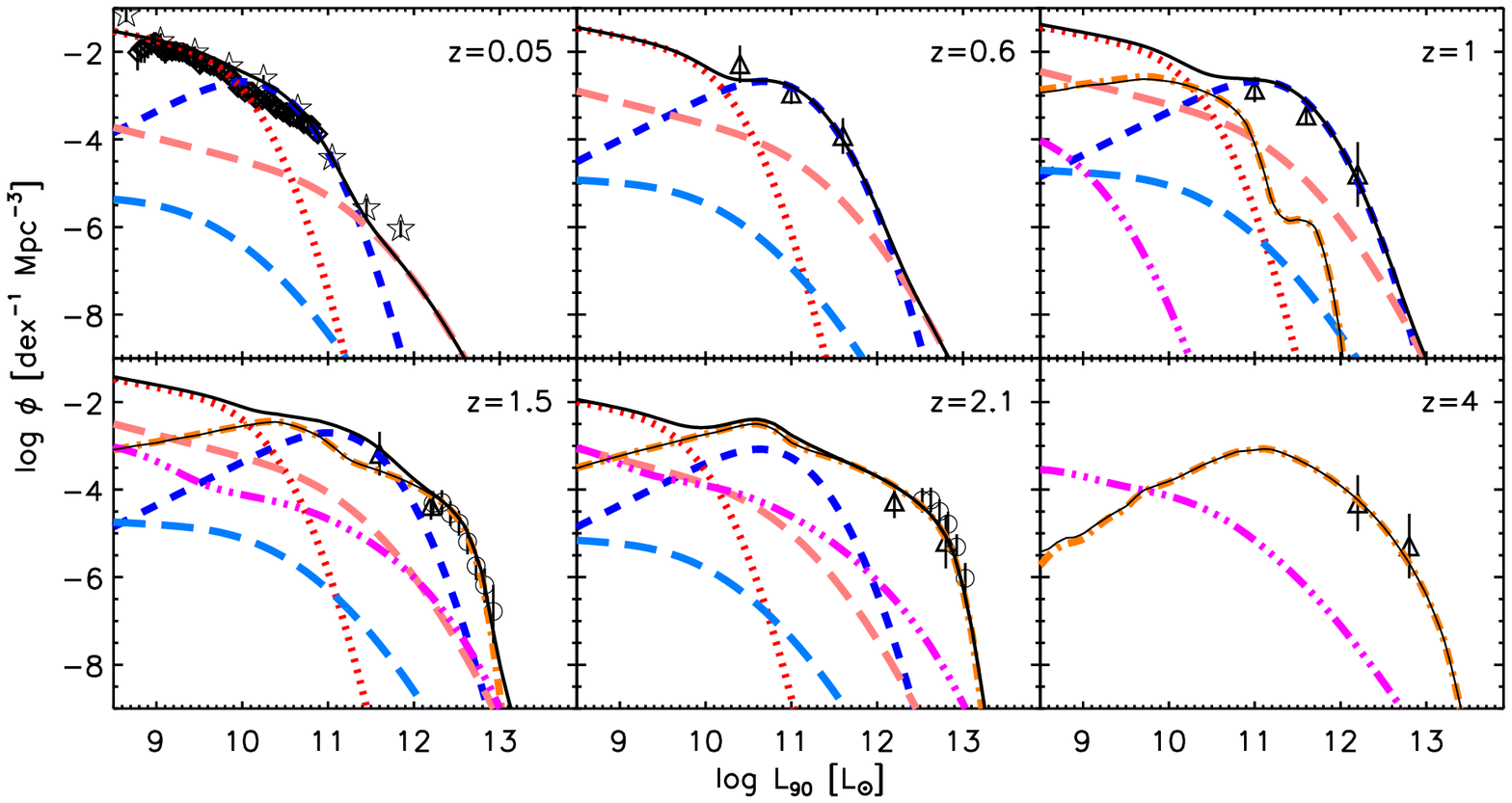}
	\caption{Comparison between model and observed $90\,\mu$m global (galaxies plus AGNs) luminosity function at several redshifts. The lines have the same definition as in Fig.~\ref{fig:LF_IR}. Data are from \citet[asterisks, 100 $\mu$m]{SoiferNeugebauer1991}, \citet[triangles]{Gruppioni2010}, \citet[diamonds]{Sedgwick2011}, and \citet[open circles, 100 $\mu$m]{Lapi2011}.  
}\label{fig:LFnu_90}
\end{center}
\end{figure*}







\clearpage

\begin{figure*} 
\begin{center}
	\includegraphics[width=\textwidth]{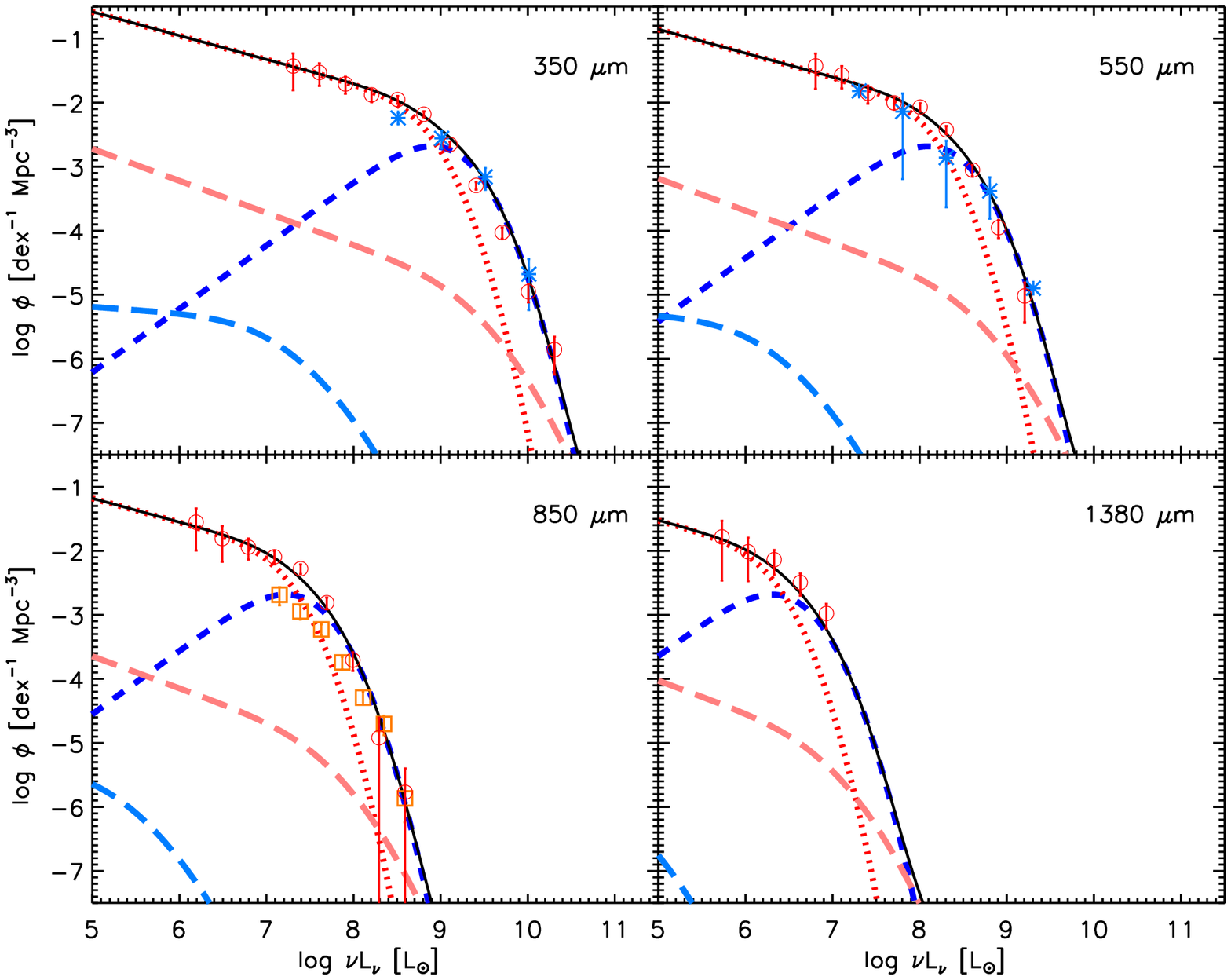}
\caption{Local luminosity functions at (sub-)mm wavelengths. As in the other figures the short-dashed blue lines refer to ``warm'' galaxies, the dotted red lines to ``cold'' galaxies, the long-dashed pink lines to type-2 AGNs and the long-dashed light-blue lines to type-1 AGNs. Data are from \citet[orange open squares]{Dunne2000}, \citet[light-blue stars]{Vaccari2010}, and \citet[red open circles]{Negrello2012}. }\label{fig:LFnu_local_submm}
\end{center}
\end{figure*}

\clearpage

\begin{figure*} 
\begin{center}
	\includegraphics[width=\textwidth]{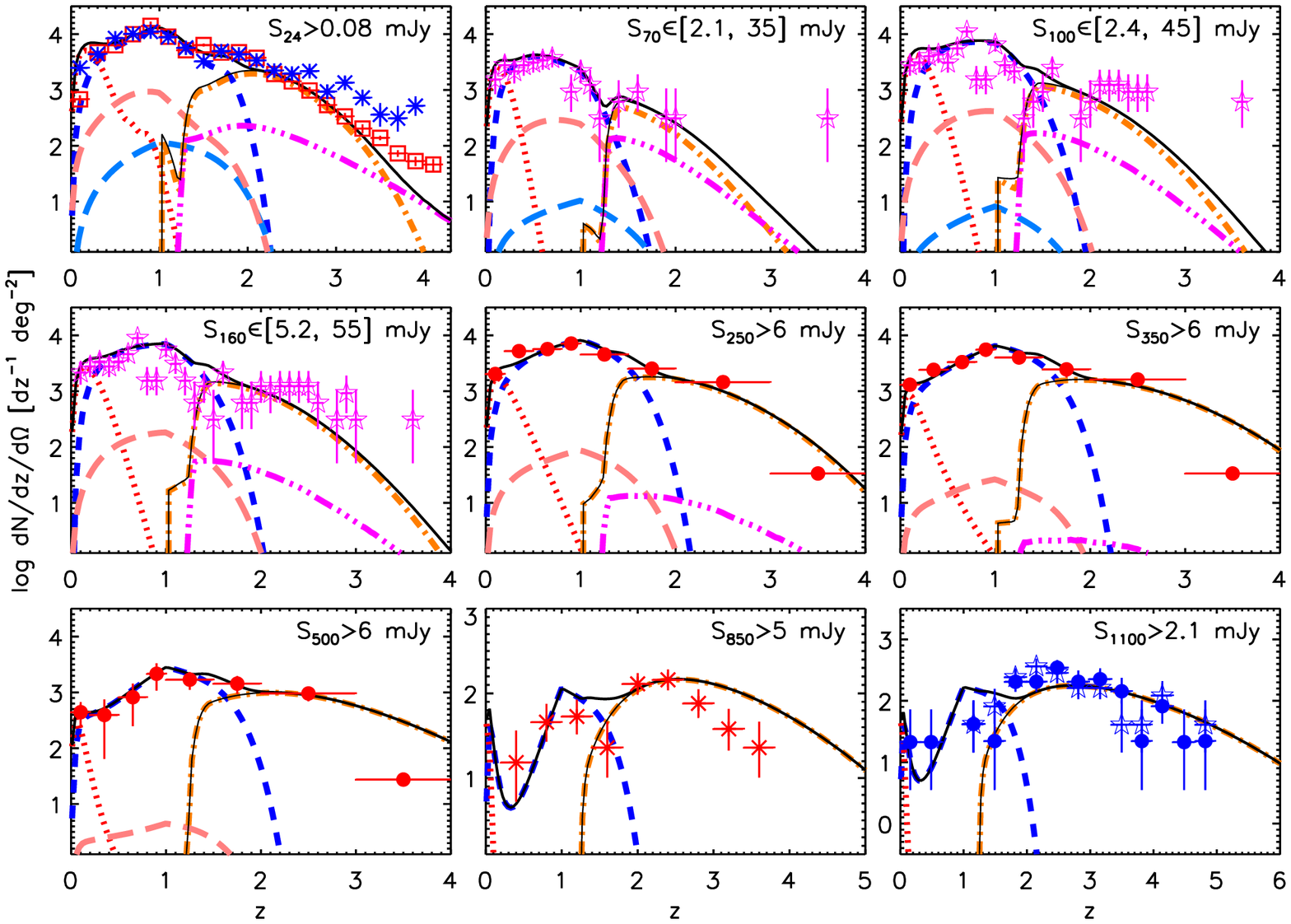}
\caption{Comparison between model and observed redshift distributions at several wavelengths and for several flux density limits. The lines have the same definition as in Fig.~\ref{fig:LF_IR}. Data are from \citet[red open squares, 24 $\mu$m]{LeFloch2009}, \citet[blue stars, 24 $\mu$m]{Rodighiero2010}, \citet[magenta open asterisks, 70, 100, and 160 $\mu$m]{Berta2011}, \citet[red filled circles, 250, 350, and 500 $\mu$m]{Bethermin2012b}, \citet[red stars, 850 $\mu$m]{Chapman2005}, and \citet[blue open asterisks based on the optical photo-$z$ and blue filled circles based on millimetric photo-$z$]{Yun2012}. Note that a substantial fraction of sources have only photometric redshifts and only few $z>2$ redshifts are spectroscopic. Photometric redshift errors tend to moderate the decline of the distributions at high-$z$; thus the observed distributions may be overestimated at the highest redshifts (see Section \ref{sect:results_lf_rd}). The dip at $z\simeq 1.5$ in the observed redshift distribution of sources with $S_{850\,\mu\rm m}> 5\,$mJy is due to the `redshift desert', i.e. to the lack of strong spectral features within the observational window and the fast decline at $z>2.5$ is due to the lack of radio identifications \citep{Chapman2005}. The dip around $z\simeq 1.5$ in the redshift distributions yielded by the model signals the transition from the phenomenological approach adopted for low-$z$ sources to the physical approach for high-$z$ proto-spheroidal galaxies and associated AGNs. Such artificial discontinuity is a weakness of the model that needs to be cleared by further work.}\label{fig:zdistr}
\end{center}
\end{figure*}

\clearpage

\begin{figure*} 
\begin{center}
	\includegraphics[width=\textwidth]{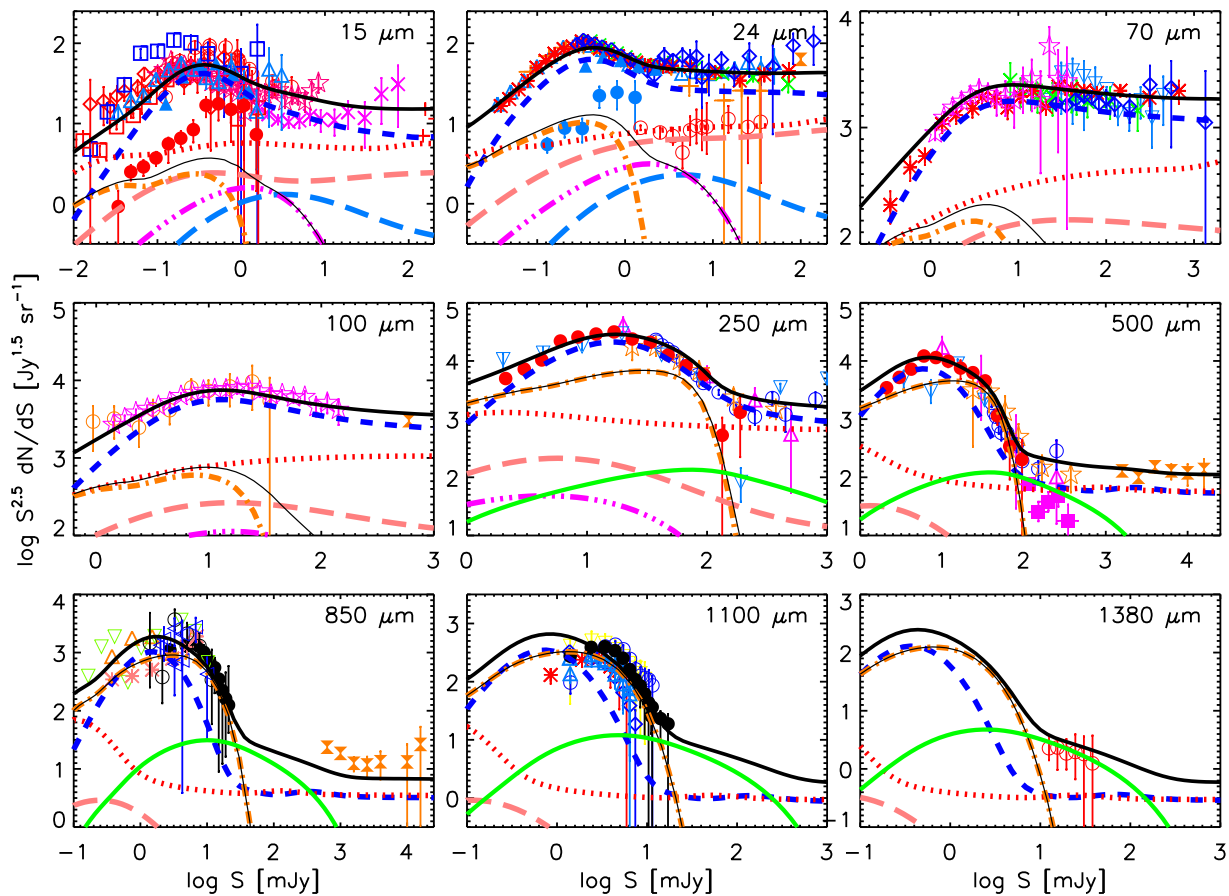}
\caption{Euclidean normalized differential number counts at wavelengths from $15\,\mu$m to $1380\,\mu$m. The thick solid lines are the sum of contributions from: ``cold'' late-type galaxies (dotted red lines), ``warm'' (starburst) late-type galaxies (dashed blue lines), type-1 AGNs (long-dashed light-blue lines), type-2 AGNs (long-dashed pink lines), stellar component of proto-spheroids (dot-dashed orange lines), AGN component of proto-spheroids (triple-dot-dashed magenta lines), strongly lensed ($\mu\ge 2$) proto-spheroids (solid green lines; only significant at $\lambda \ge 250\,\mu$m). The thin solid black lines show the counts of unlensed proto-spheroids, including both the stellar and the AGN components; at $\lambda \ge 250\,\mu$m these counts essentially coincide with the counts of the stellar component only. The filled red circles in the $15\,\mu$m panel refer to AGNs only. The filled blue circles and the open red circles in the $24\,\mu$m panel refer to AGNs only and come from \citet{Treister2006}  and from \citet{Brown2006}, respectively. The purple filled squares in the $500\,\mu$m panel show the estimated counts of strongly lensed galaxies \citep{Lapi2012}. The bright counts at 1.38 mm are also generally interpreted as due to strongly lensed galaxies \citep{Vieira2010,Greve2012}. References for all the data points are given in Table~\protect\ref{nc_references}. The model provides a physical explanation of the sudden steepening of the (sub-)mm counts: it is due to the appearance of proto-spheroidal galaxies that show up primarily at $z \gsim 1.5$, being mostly in passive evolution at lower redshifts.
}\label{fig:nc_dnc}
\end{center}
\end{figure*}
\clearpage

\begin{figure*} 
\begin{center}
	\includegraphics[width=\textwidth]{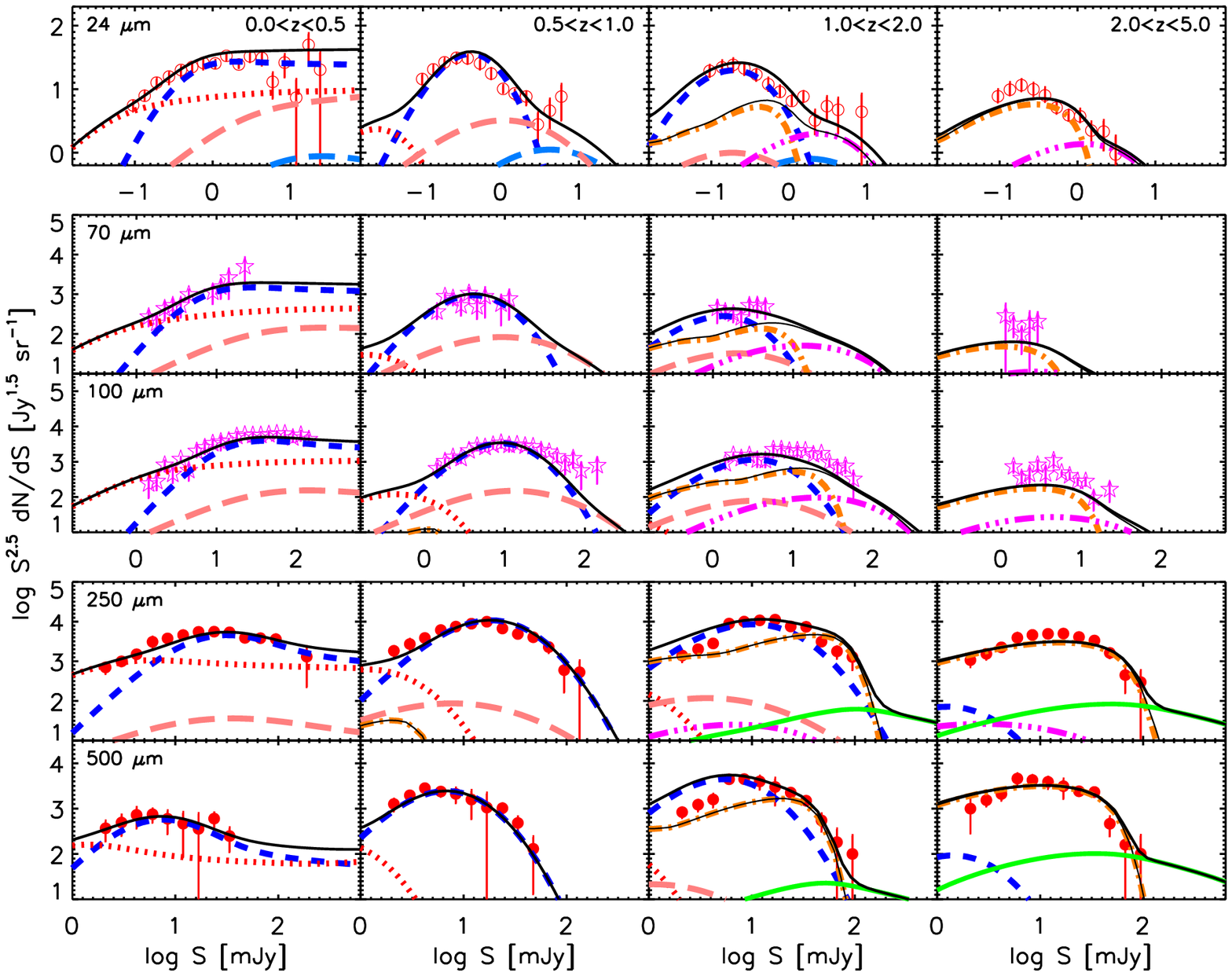}
\caption{Euclidean normalized differential number counts per redshift slices. Lines have the same meaning as in Fig.~\protect\ref{fig:nc_dnc}. Data are from \citet[red open circles, 24 $\mu$m]{LeFloch2009}, \citet[magenta open asterisks, 70 and 100 $\mu$m]{Berta2011}, and \citet[red filled circles, 250 and 500 $\mu$m]{Bethermin2012b}. }\label{fig:nc_dnc_zbin}
\end{center}
\end{figure*}

\clearpage

\begin{figure*} 
\begin{center}
	\includegraphics[width=\textwidth]{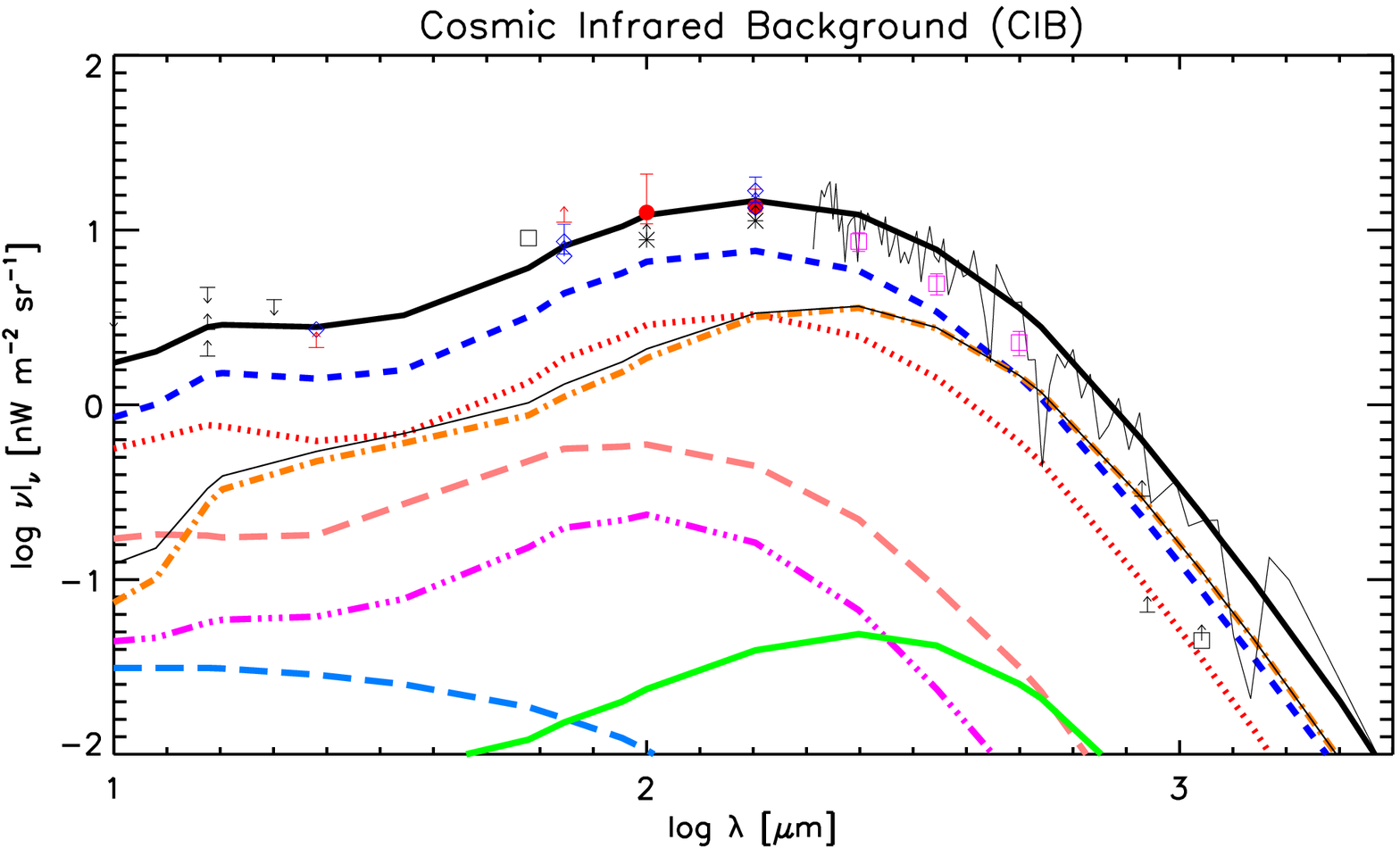}
	\caption{Contributions of the different populations to the cosmic infrared background. The lines have the same meaning as in Fig.~\ref{fig:nc_dnc}. Proto-spheroidal galaxies are the main contributors to the CIB above $\simeq 500\,\mu$m. Data points are from \citet{Renault2001}, \citet{SteckerDeJager1997}, \citet{Lagache1999}, \citet{Elbaz2002}, \citet{Miville-Deschenes2002}, \citet{Smail2002}, \citet{Papovich2004}, \citet{Dole2006}, \citet{Marsden2009}, \citet{Hopwood2010}, \citet{Greve2010}, \citet{Scott2010}, \citet{Altieri2010}, and \citet{Berta2011}. }\label{fig:CIB}
\end{center}
\end{figure*}

\clearpage


\clearpage

\begin{figure*} 
\begin{center}
\includegraphics[scale=0.3]{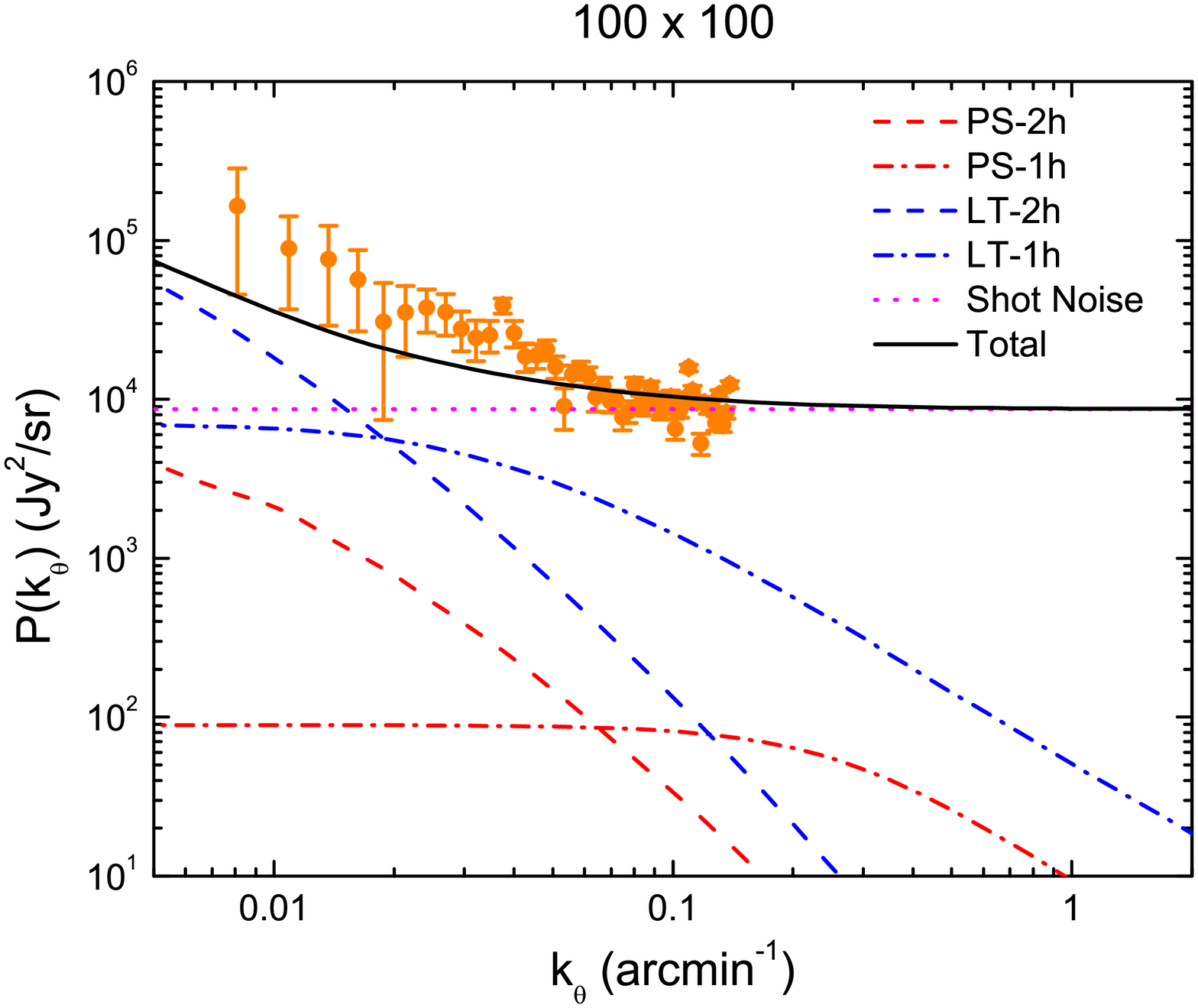}
\includegraphics[scale=0.3]{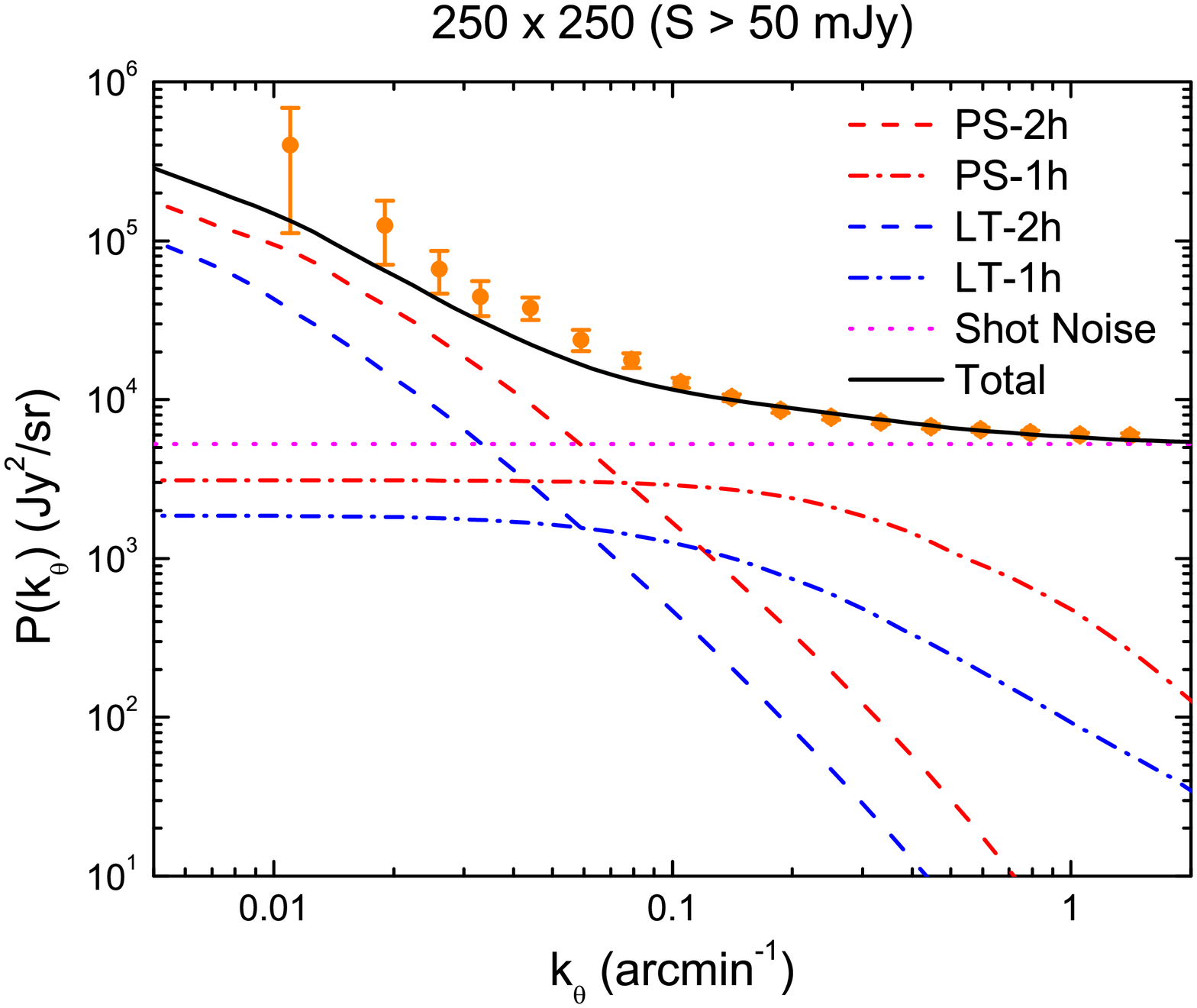}
\includegraphics[scale=0.3]{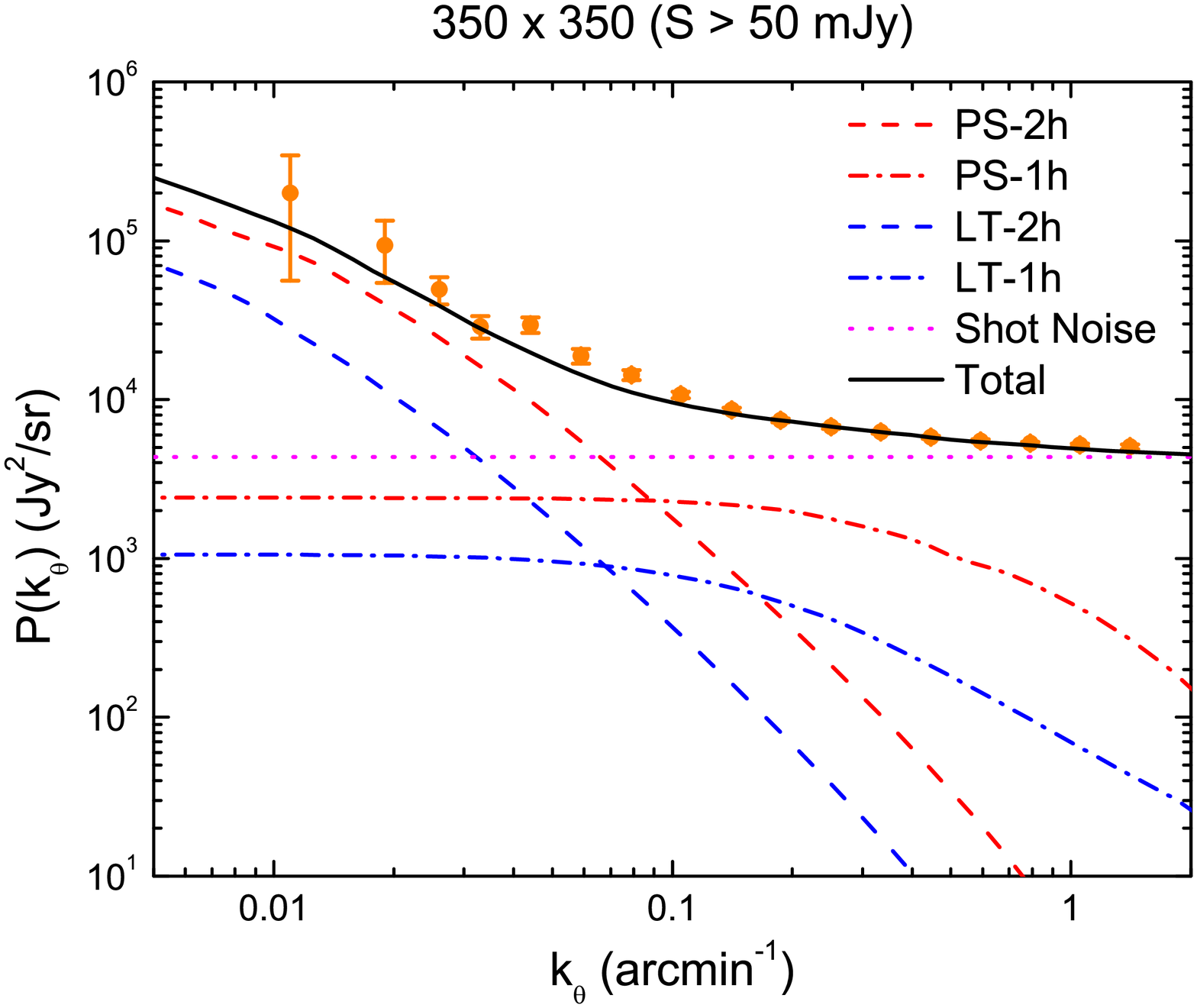}
\includegraphics[scale=0.3]{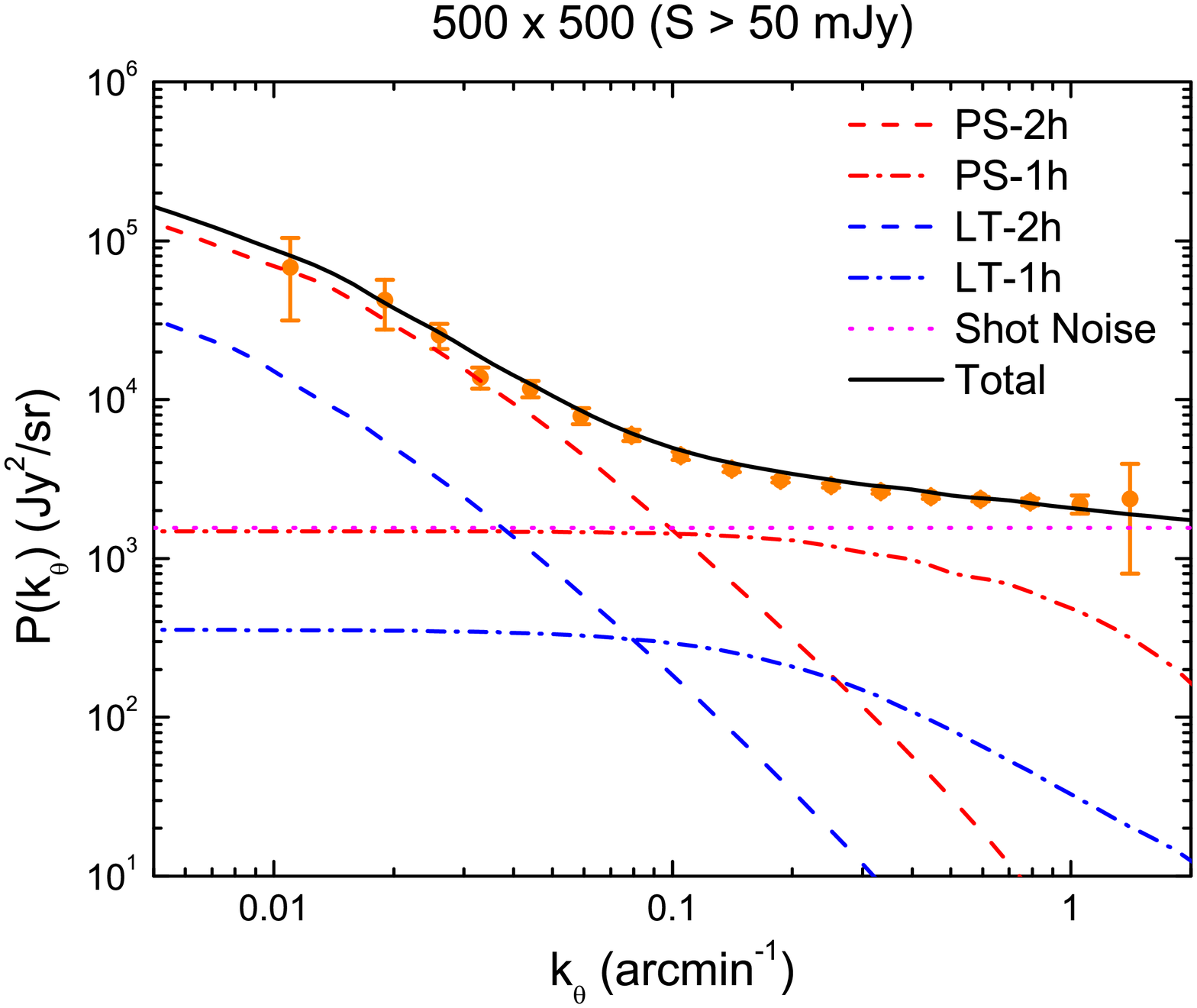}
\caption{CIB angular power spectra at far-IR/sub-mm wavelengths. The $100\,\mu$m data are from \citet{Penin2012}, those at longer wavelengths are from \citet{Viero2012}. The lines show the contributions of the 1-halo and 2-halo terms for the two populations considered here: late-type (LT) ``warm'' plus ``cold'' galaxies and proto-spheroidal (PS) galaxies. The horizonal dotted magenta lines denote the shot noise level. At $\lambda \ge 250\,\mu$m the signal is dominated by proto-spheroidal galaxies while late-type galaxies take over at shorter wavelengths.
}\label{fig:pk}
\end{center}
\end{figure*}

\clearpage

\begin{figure*} 
\begin{center}
\includegraphics[scale=0.19]{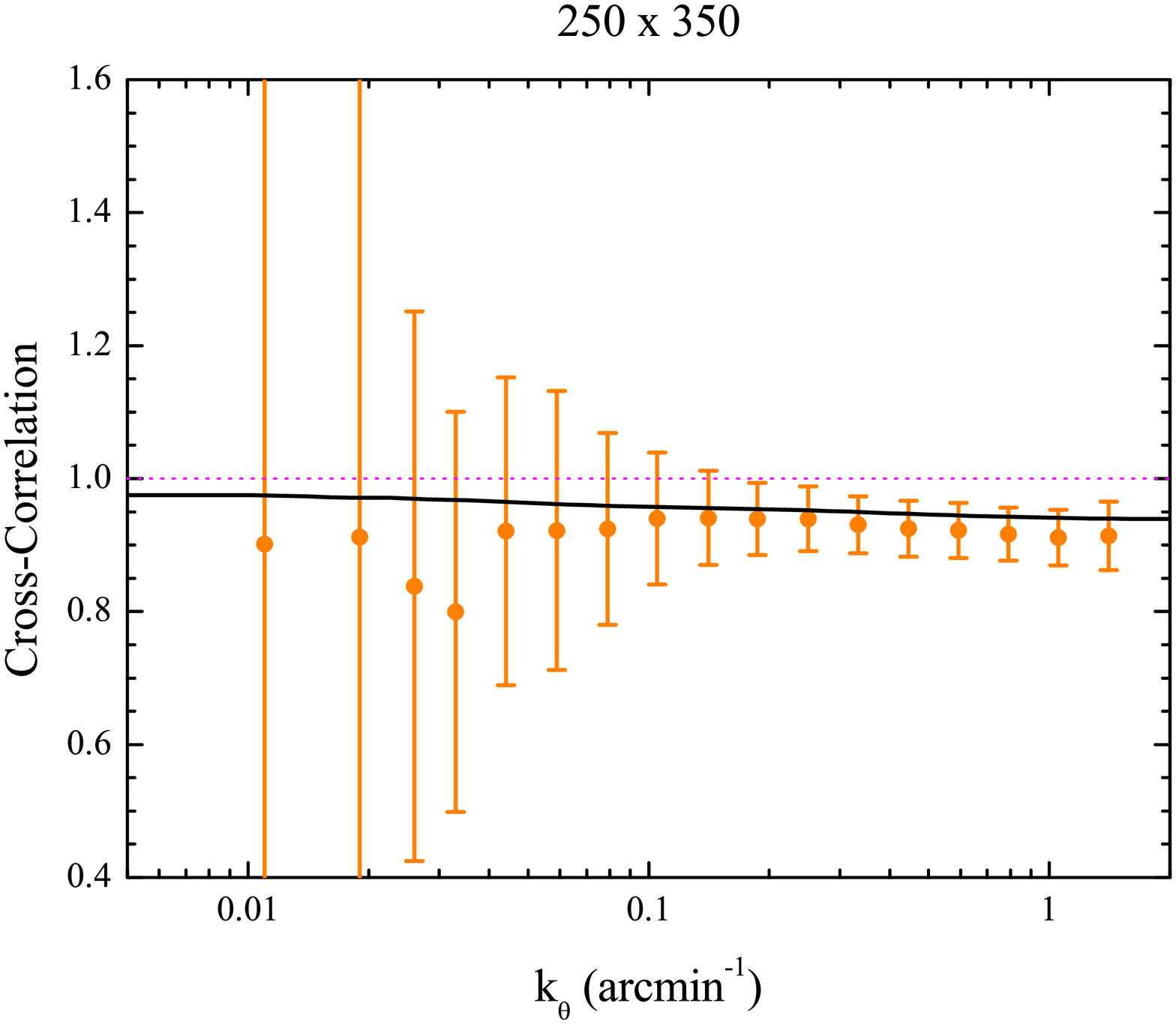}
\includegraphics[scale=0.19]{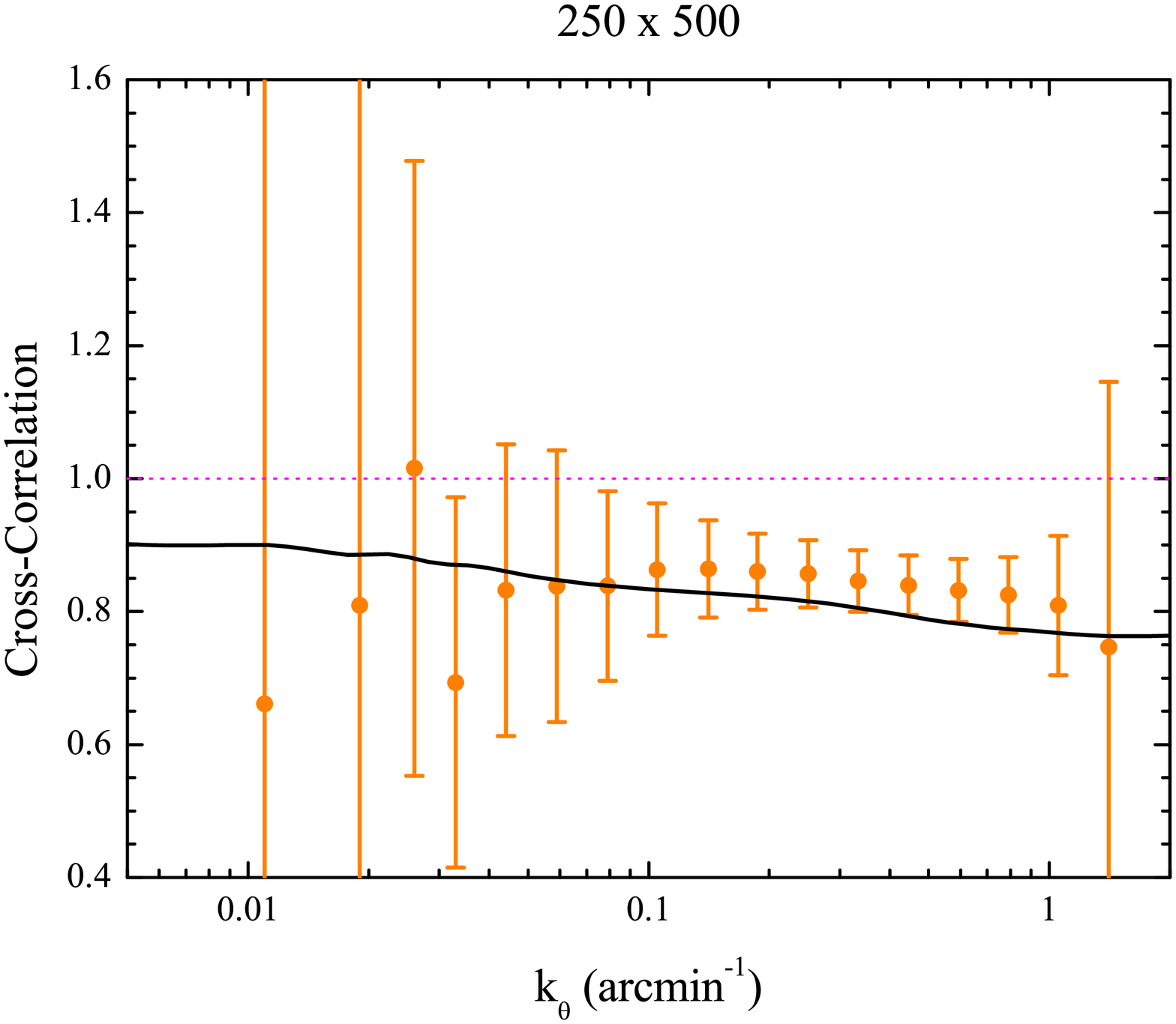}
\includegraphics[scale=0.19]{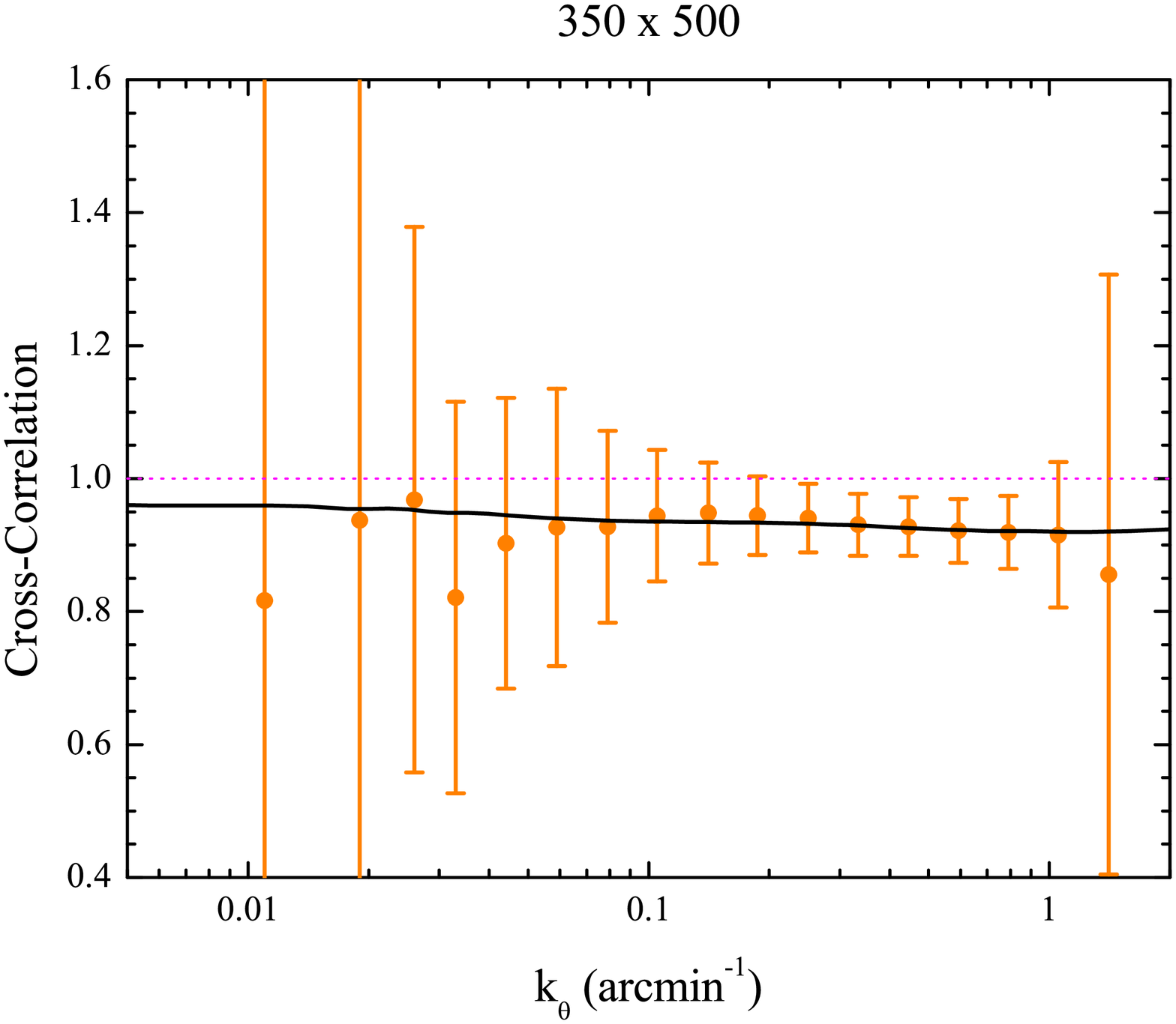}
\caption{CIB cross-frequency power spectra at sub-mm wavelengths normalized according to eq.~(14) of \citet{Viero2012}. The solid line is the result from the model. The data are from \citet{Viero2012}.
}\label{fig:norm_cross_spectra}
\end{center}
\end{figure*}

\clearpage

\begin{figure*} 
  \begin{center}
   \includegraphics[scale=0.8]{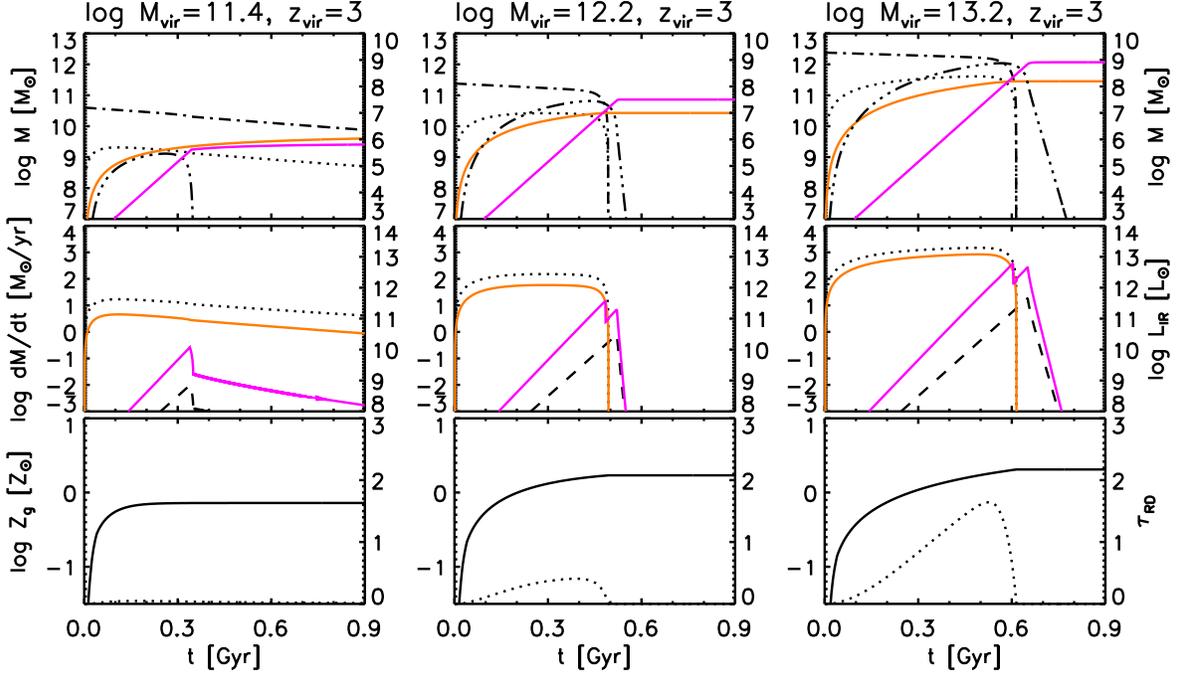}
   \caption{Evolution with galactic age of properties of the stellar and of the AGN component of proto-spheroidal galaxies virialized at $z_{\rm vir}=3$ for three choices of the virial (mostly dark matter) mass: $\log(M_{\rm vir}/M_\odot)=11.4$ (left-hand column), 12.2 (central column) and 13.2 (right-hand column). In the first row, the left y-axis scale refers to masses related to the stellar component [infalling hot gas mass (dot-dashed line), cold gas mass (dotted line), stellar mass ($M_{\star}$, solid orange line)] while the right-hand scale refers to quantities related to the AGN component [reservoir mass (triple-dot-dashed line) and black-hole mass ($M_\bullet$, solid magenta line)]. In the second row the left-hand scale refers to the SFR (dotted line) and to the BH accretion rate (dashed black line), while the right-hand scale refer to the IR (8--$1000\,\mu$m) luminosity of the stellar (solid orange line) and of the AGN (solid magenta line) component. In the third row the left-hand scale refers to the gas metallicity (solid line) while the right-hand scale refers to the optical depth of individual gas clouds (dotted line).}\label{fig:sph_evol_mlz3}
  \end{center}
\end{figure*}

\clearpage

\begin{table*}
\caption{Parameters for low-z AGNs and for ``warm'' and ``cold'' galaxy populations. The parameters of the AGN luminosity functions refer to $12\,\mu$m ($\nu L_\nu$) while those for galaxies refer to IR (8--$1000\,\mu$m) luminosities. Values without error were kept fixed.}\label{tab:parameters}
\begin{center}
\scalebox{1.0}{
	\begin{tabular}{l c c c c c c }
	\hline
	\hline
	 & AGN~1 (12 $\mu$m) & AGN~2 (12 $\mu$m) & Warm (IR) & Cold (IR) \\
	\hline
	$\log (\Phi^*_0/\hbox{Mpc}^{-3})$ & -5.409 $\pm$ 0.098 & -4.770 $\pm$ 0.122 & -2.538 $\pm$ 0.051 & -1.929 $\pm$ 0.112 \\
	$\log (L^*_0/L_\odot)$    &  9.561 $\pm$ 0.084 & 10.013 $\pm$ 0.093 & 10.002 $\pm$ 0.076 & 9.825 $\pm$ 0.087 \\
	$\alpha$   & 1.1  & 1.5 & 0.01 & 1.372 $\pm$ 0.121  \\
	$\sigma$   & 0.627 $\pm$ 0.017 & 0.568 $\pm$ 0.021 & 0.328 $\pm$ 0.014 & 0.3 \\
	$\alpha_\Phi$ & 2.014 $\pm$ 0.400 &  4.499 $\pm$ 0.317 & 0.060 $\pm$ 0.200 & 0.0 \\
	$\alpha_L$ & 2.829 $\pm$ 0.297 &  0.0 & 3.625 $\pm$ 0.097 & 1.0 \\
	$z_{\rm break}$  & 1.0 & 1.0  & 1.0 & 1.0 \\
	$z_{\rm cutoff}$ & 2.0 & 2.0 & 2.0 & 2.0 \\
	\hline
	\end{tabular}
}
\end{center}
\end{table*}

\clearpage

\begin{table*}
\caption{Parameters of the physical model for the evolution of proto-spheroidal galaxies and associated AGNs. The values of the first eight parameters used here are somewhat different from those used in previous papers, but still well within the plausible ranges listed in column 3 and discussed in the references given in the footnotes. }\label{tab:parameters_proto}
\begin{center}
\scalebox{0.90}{
	\begin{tabular}{l c c l }
	\hline
	\hline
	Parameter & Value & Plausible range & Description \\
	\hline
	$\tau^0_{\rm RD}$ & 3.0 & 1 - 10$^{\rm a}$ & Normalization of optical depth of gas cloud [eq.~(\ref{eq:optical_depth})]\\
	$\epsilon$ & 0.10 & 0.06 - 0.42 & Black hole accretion radiative efficiency [eq.~(\ref{eq:Lbh})]\\
	$\lambda_{\rm Edd}$ & 1 - 4& $\lesssim$ a few$^{\rm b}$  & Redshift dependent Eddington ratio [eq.~(\ref{eq:Edd_ratio})]\\
	$\epsilon_{\rm QSO}$ & 3.0 & 1 - 10$^{\rm a}$ & Strength of QSO feedback [eq.~(\ref{eq:L_QSO})]\\
	$k_{\star, \rm IR}$ & 3.1 & 2 - 4$^{\rm c}$ & Conversion factor from of SFR to IR luminosity [eq.~(\ref{eq:SFR-IR})]\\
	$\sigma_*$ & 0.10 & $\lesssim$ 0.5 & Dispersion of mean stellar luminosity [eq.~(\ref{LF_scatter2})]\\
	$\sigma_\bullet$ & 0.35 & $\lesssim$ 0.5$^{\rm b}$ & Dispersion of mean AGN luminosity [eq.~(\ref{LF_scatter2})]\\
	$f_{\rm gas,crit}$ & 0.03 & $\lesssim$ 0.165 & Gas mass fraction at transition \\
                       &      &                  & from obscured to unobscured AGNs [\S\,\ref{sect:protoLF})]\\
\hline
	$\epsilon_{\rm SN}$ & 0.05 & 0.01 - 0.1$^{\rm d}$ & Strength of SN feedback [eq.~(\ref{eq:SN})]\\
	$\alpha_{\rm RD}$ & 2.5 & 1 - 10$^{\rm e}$ & Strength of radiation drag [eq.~(\ref{eq:RD})]\\
	\hline
	\end{tabular}
}\\
\begin{flushleft}
	$^{\rm a}$\,\citet{Granato2004}; $^{\rm b}$\,\citet{Lapi2006}; $^{\rm c}$\,\citet{Lapi2011}; $^{\rm d}$\,\citet{Shankar2006}; $^{\rm e}$\,Lapi et al. (in preparation).\\
\end{flushleft}
\end{center}
\end{table*}


\clearpage

\begin{table}[htb]
\caption{References for data on number counts (see Fig.~\protect\ref{fig:nc_dnc})}
\begin{center}
\scalebox{0.75}{
	\begin{tabular}{l c c r}
	\hline
	\hline
	Wavelength ($\mu$m) & Instrument & Field & Reference \\
	\hline
	15, 24 & AKARI/IRC & NEP-deep & \citet{Takagi2012} \\
	15 & AKARI/IRC & de-lensed Abell 2218 & \citet{Hopwood2010} \\
	15 & AKARI/IRC & NEP-deep+wide & \citet{Pearson2010} \\
	15 & AKARI/IRC & CDFS & \citet{Burgarella2009} \\
	15 & AKARI/IRC & NEP-deep & \citet{Wada2007} \\
	15 & ISO/ISOCAM & ELAIS-S & \citet{Gruppioni2002} \\
	15 & ISO/ISOCAM & ISOCAM deep surveys & \citet{Elbaz1999} \\
	16 & Spitzer/IRS & GOODS-N+S & \citet{Teplitz2011} \\
	24, 70 & Spitzer/MIPS & ADF-S & \citet{Clements2011} \\
	24, 70 & Spitzer/MIPS & Spitzer legacy fields & \citet{Bethermin2010} \\
	24 & Spitzer/MIPS & SWIRE fields & \citet{Shupe2008} \\
	24 & Spitzer/MIPS & NDWFS Bootes & \citet{Brown2006} \\
	24 & Spitzer/MIPS & GOODS-N & \citet{Treister2006} \\
	24 & Spitzer/MIPS & Deep Spitzer fields & \citet{Papovich2004} \\
	70, 100 & Herschel/PACS & GOODS, LH, COSMOS & \citet{Berta2011} \\
	70 & Spitzer/MIPS & xFLS & \citet{Frayer2006} \\
	70 & Spitzer/MIPS & Bootes, Marano, CDF-S & \citet{Dole2004} \\
	100 & Herschel/PACS & Abell 2218 & \citet{Altieri2010} \\
	250, 500 & Herschel/SPIRE & HerMES & \citet{Bethermin2012b} \\
	250, 500 & Herschel/SPIRE & H-ATLAS & \citet{Clements2010} \\
	250, 500 & Herschel/SPIRE & HerMES & \citet{Oliver2010} \\
	250, 500 & Herschel/SPIRE & HerMES P(D) & \citet{Glenn2010} \\
	250, 500 & BLAST & BGS P(D) & \citet{Patanchon2009} \\
	500 & Herschel/SPIRE & H-ATLAS & \citet{Lapi2012} \\
	550, 850 & Planck & Planck all-sky survey & \citet{PlanckCollaboration2012} \\
	850 & SCUBA & Clusters & \citet{Noble2012} \\
	850 & SCUBA & Abell 370 & \citet{Chen2011} \\
	850 & SCUBA & Clusters & \citet{Zemcov2010} \\
	850 & SCUBA & Clusters \& NTT-DF & \citet{Knudsen2008} \\
	850 & SCUBA & SHADES & \citet{Coppin2006} \\
	850 & SCUBA & Clusters & \citet{Smail2002} \\
	870 & APEX/LABOCA & Clusters & \citet{Johansson2011} \\
	1100 & ASTE/AzTEC & AzTEC blank-field survey & \citet{Scott2012} \\
	1100 & ASTE/AzTEC & COSMOS & \citet{Aretxaga2011} \\
	1100 & ASTE/AzTEC & ADF-S, SXDF \& SSA22 & \citet{Hatsukade2011} \\
	1100 & ASTE/AzTEC & GOODS-S & \citet{Scott2010} \\
	1100 & JCMT/AzTEC & SHADES & \citet{Austermann2010} \\
	1100 & JCMT/AzTEC & COSMOS & \citet{Austermann2009} \\
	1400 & SPT & SPT survey & \citet{Vieira2010} \\
	\hline
	\end{tabular}
}
\end{center}
\label{nc_references}
\end{table}

\end{document}